%% file: RPV_SUSY_EWK_2023.tex
\documentclass{ws-ijmpa}

\usepackage{amsthm}
\usepackage{amsfonts}
\usepackage{xspace}

\usepackage{graphicx}
\usepackage{multirow}
\usepackage{subfigure}
\usepackage{lineno}
\usepackage{hyperref}
\usepackage[normalem]{ulem}
\hypersetup{
	colorlinks=true,
	linkcolor=blue,
	citecolor=blue,
	urlcolor=blue
}
\newcommand{\customUL}{\bgroup\markoverwith{\textcolor{blue}{\rule[-0.3ex]{2pt}{0.6pt}}}\ULon}
\newcommand{\link}[2]{\textcolor{blue}{\href{#1}{\customUL{#2}}}}

\usepackage[super,compress]{cite}
\usepackage{graphicx}

\newcommand{\AtlasCoordFootnote}{%
	ATLAS uses a right-handed coordinate system.
	The pseudorapidity is defined in terms of the polar angle \(\theta\) as \(\eta = -\ln \tan(\theta/2)\).
	Angular distance is measured in units of \(\Delta R \equiv \sqrt{(\Delta\eta)^{2} + (\Delta\phi)^{2}}\).}

\input{mydefs.tex}

\begin{document}
\markboth{Otilia Ducu}{ATLAS searches for higgsinos with R-parity violating couplings in events with leptons}

%
%

\title{ATLAS searches for higgsinos with R-parity violating couplings in events with leptons
}

\author{Otilia Anamaria Ducu\\ (on behalf of the ATLAS Collaboration)
}

\address{Horia Hulubei National Institute of Physics and Nuclear Engineering (IFIN-HH)\\
Magurele, Ilfov, Romania, 077125\\
otilia.ducu@gmail.com}

\maketitle


\begin{abstract}
This document presents two searches for Supersymmetry through the direct production of pairs of higgsinos decaying into final states with leptons and ($b$-) jets. The analyses are performed using 139~fb$^{-1}$ of the 13~TeV proton-proton collision data collected with the ATLAS detector.
The methods used to estimate the Standard Model and detector backgrounds are discussed, as well as their shortcomings. Finally, results in selected signal regions, and some exclusion limits, are presented, illustrating the significant improvement over the previous exclusion limits.

\textit{Document based on a presentation at the XI International Conference on New Frontiers in Physics (ICNFP 2022).}

\end{abstract}

\ccode{PACS numbers: 11.30.Pb, 12.60.Jv}


%
%

\section{Introduction}	

Supersymmetry~\cite{Martin:1997ns} (SUSY) remains one of the preferred extensions of the Standard Model (SM) of particle physics, despite the absence of experimental evidence in the LHC~\cite{Evans:2008zzb} Run 1 and Run 2 data. 
This is mainly because it can solve the SM gauge hierarchy problem without a large fine tuning of fundamental parameters, by predicting for each SM particle a super-partner with a half spin difference.
Moreover, weakly interacting particles that are good dark matter candidates are present in the list of SUSY particles: the lightest neutralino ($\ninoone$) or the gravitino.

\begin{figure}[!tb]
	\begin{center}
		\subfigure[]{\label{fig:diag_biRPV1}
			\includegraphics[width=0.22\columnwidth]{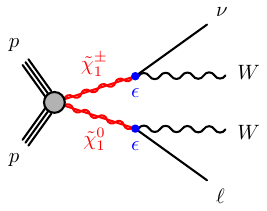}
		}\hspace{-0.05cm}
		\subfigure[]{\label{fig:diag_biRPV2}
			\includegraphics[width=0.22\columnwidth]{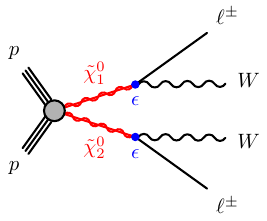}
		}\hspace{-0.05cm}
		\subfigure[]{\label{fig:diag_RPV1}
			\includegraphics[width=0.22\columnwidth]{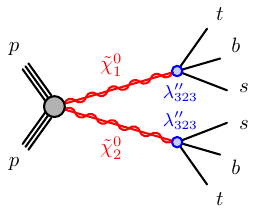}
		}\hspace{-0.05cm}
		\subfigure[]{\label{fig:diag_RPV2}
			\includegraphics[width=0.22\columnwidth]{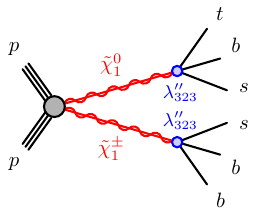}
		}\vspace{-0.3cm}	
	\end{center}
	\caption{
		Representative diagrams illustrating the production and subsequent RPV decays of the higgsinos. Reused with permission from Ref.~\citen{ATLAS-CONF-2022-057,ATLAS:2021fbt}.
	}
	\label{fig:diag_biRPV_RPV}
\end{figure}

Only the weak production of higgsinos is studied in the searches discussed in this document. Two scenarios with $R$-parity violated~\cite{Dreiner:1997uz} (RPV) are considered:
\begin{itemize}
	\item[1)] Higgsino {bRPV~\cite{ATLAS-CONF-2022-057}}: the $R$-parity violation is obtained through lepton-number violation. Here, bilinear terms were introduced to the superpotential (bRPV). To ensure higgsino decays to light leptons, thus to suppress decays to tau leptons, tan$\beta$ parameter is set to 5. The higgsinos are nearly degenerate, with a mass splitting of approximately 2~GeV. The considered production modes are $\chinoonepm \ninoone$, $\chinoonepm\ninotwo$ and $\ninoone\ninotwo$. All higgsino possible bRPV decays are allowed in the model, with the dominant decays being: $\chinoonepm\to W^\pm\nu_{\mu}$ and $\susy{\chi}^0_{1,2}\to W^\pm\ell^{\mp}, W^\pm\tau^{\mp}$. Some representative diagrams are shown in Figures~\ref{fig:diag_biRPV1} and~\ref{fig:diag_biRPV2}.
	\item[2)] Higgsino UDD {RPV~\cite{ATLAS-CONF-2022-057,ATLAS:2021fbt}}: the $R$-parity violation is obtained through baryon number violation, and the $UDD$-type coupling $\lambda''_{323}$ is chosen to be non-vanishing. To ensure only prompt decays for $\ninoone$ and $\ninotwo$ with masses $>180$~GeV, $\lambda''_{323}$ is set to have a value of $\mathcal{O}(10^{-3}-10^{-2})$. The $\ninoone$ sparticle is always the LSP. The $\chinoonepm$ and $\ninotwo$ sparticles are assumed to be effectively mass degenerate with the LSP, and all other electroweakinos are assumed to be decoupled and not considered in the model. Both \ninotwo\ and \ninoone\ decay in the same way, $\susy{\chi}_{1,2}^0 \to tbs$ with a 100\% BR; \chinoonepm\ decays to $sbb$ with a 100\% BR. For this simplified model, the considered diagrams are shown in Figures~\ref{fig:diag_RPV1} and~\ref{fig:diag_RPV2}.
	\item Signal event samples for these models are generated using MadGraph5 aMC@NLO interfaced to Pythia8 for the modelling of the parton showering, hadronisation and underlying {event~\cite{ATLAS-CONF-2022-057,ATLAS:2021fbt}}. 
\end{itemize}

As can be seen in Figure~\ref{fig:diag_biRPV_RPV}, the considered RPV models have either top quarks or $W$ bosons in the decay chains, ensuring the possibility of leptonic final states. The searches presented in this document are  performed only with events containing one lepton, two same electric charge leptons or three leptons. These, as well as the zero lepton channels, are being extensively considered by both ATLAS and CMS experiments, and the obtained results were published in numerous {articles~\cite{SUSY_ATLAS,SUSY_CMS}}.

Even if the signatures including leptons are characterized by small branching ratios, they are well motivated as the level of SM background is quite low. In this document, the focus is mainly on the work done in the channels with
two same-charge or three leptons, as the background estimation is more challenging. However, selected results in the one lepton channel are presented for the UDD RPV model illustrated in Figure~\ref{fig:diag_RPV2}.

For these studies, the LHC Run 2 data set of proton-proton collisions at $\sqrt{s}=13$~TeV recorded by the ATLAS {detector~\cite{Atlas_Detector}}, and corresponding to an integrated luminosity of 139~fb$^{-1}$ is used.
The ATLAS detector is a multipurpose particle detector with a forward--backward symmetric cylindrical geometry and a near \(4\pi\) coverage in solid angle\footnote{\AtlasCoordFootnote}.
It consists of an inner tracking detector surrounded by a thin superconducting solenoid providing a 2~T axial magnetic field, electromagnetic and hadron calorimeters, and a muon spectrometer.


Interesting events are selected by the first-level trigger system implemented in custom hardware,
followed by selections made by algorithms implemented in software in the high-level {trigger~\cite{ATLAS:2016wtr}}. 
For the analysis performed only with two same-charge or three leptons, events are selected with the lowest unprescaled di-lepton triggers~\cite{ATLAS:2019dpa,ATLAS:2020gty} and missing transverse energy (\met) based {triggers~\cite{ATLAS:2020atr}}. 
For the analysis including also the one lepton event topology, events are selected with the lowest unprescaled single lepton triggers. 
The employed trigger selection is ensuring a maximum, and a rather constant with respect to the lepton transverse momentum (\pT) trigger efficiency.

An extensive software suite~\cite{ATL-SOFT-PUB-2021-001} was used in data simulation, in the reconstruction and analysis of real and simulated data, in detector operations, and in the trigger and data acquisition systems of the experiment.

\section{Strategy to look for RPV SUSY}
The following strategy is used to search for higgsino signals. 
At first, dedicated discovery and exclusion signal regions are optimized for each RPV model. This is done via a complete scan of different sets of cuts on the background discriminant kinematic variables representative for the final state. During the optimization, the signal sensitivity is evaluated through its potential discovery significance. Only Monte-Carlo (MC) simulations are used, and to account for uncertainties on the background prediction a 30\% systematic uncertainty is considered.
To ensure the best sensitivity, more than one signal region per RPV model is defined.
The final signal regions can overlap.

Once the final signal regions are selected, the SM and detector background sources are identified and further estimated as best as possible.
An extensive validation is also performed, to ensure the correctness of the employed methods.
These measurements and checks are done with the help of control and validations regions that are defined for the main background sources.

A simultaneous fit method is used to compute the final signal and background uncertainty estimates, and to statistically test the compatibility between data and background estimation in the signal regions. 
The method is based on a profile-likelihood ratio test and is implemented in the Histfitter {tool~\cite{Baak:2014wma}}.
Three different fits are performed to get the final results. A background-only fit~\cite{Baak:2014wma} is used to estimate the total background in the signal and validation regions, without assumptions on the signal model.
These background predictions are independent of the observed number of data events in each signal and validation region. 
A background-only fit is used to also estimate the normalization factor (and its uncertainty) of the $WZ$+jets background.

A model-dependent signal fit~\cite{Baak:2014wma} is used to obtain the exclusion limits. The signal contribution is accounted for, and is given by the number of signal events estimated in the signal region(s) considered in the fit.
The fit is performed only in the exclusion signal {region(s)~\cite{ATLAS-CONF-2022-057}}, or simultaneously in the signal and control {regions~\cite{ATLAS:2021fbt}}. 
It is important to note that more than one signal or control region can be included in this fit only if they are orthogonal (thus, statistically independent).  

A model-independent signal fit~\cite{Baak:2014wma} is used to obtain the model-independent upper limits on the number of events beyond the expected number of events in each discovery signal region. 
This information is important especially for theorists, as it can be used to probe any BSM model.
As for the model-dependent signal fit, both signal and control regions are used. No assumptions are made for the signal model, and the number of signal events in the signal region is added as a parameter to the fit.

\section{Signal region definitions}

A detailed description of \met, lepton and ($b$-) jet objects used to get the results presented in this document is given in Ref.~\citen{ATLAS-CONF-2022-057,ATLAS:2021fbt}. 
Baseline (signal) leptons collections are defined using loose (tight) lepton identification (plus isolation) criteria, as well as requirements on the transversal and longitudinal impact parameters. 
The leading and sub-leading (third-leading) leptons are required to have $\pT > 15$~GeV (10~GeV).
The selection requirements are a compromise between a reasonably-efficient identification of prompt leptons and a good rejection of background leptons. 
For example, this efficiency can reach values higher than 90\% for leptons with $\pT>100$~GeV, in a $Z\to \ell^\pm \ell^\mp$ event selection.
Jet candidates are reconstructed using the anti-$k_t$ algorithm, with radius parameter $R=0.4$ and particle-flow objects as {inputs~\cite{ATLAS:2017ghe}}. For the $b$-jet identification, a  DL1r $b$-tagging {algorithm~\cite{ATLAS:2019bwq,ATLAS:bjets}} is used, with an average identification efficiency of 70\% in a \ttbar\ event selection. Only ($b$-) jets with $\pT > 20$~GeV are used further. 

In addition to the counting of the objects discussed above, the discriminating variables presented below are used to define the signal regions.
\begin{itemize}
	\item \met, that has moderate values in signal events as it comes only from neutrinos.
	\item The stransverse mass, $\mttwo$, an event variable that is correlated to the masses of an unseen pair of particles that are presumed to have decayed semi-invisibly into particles that are seen. It is defined as a function of the momenta of two visible particles and the \met in the event. 
	\item The inclusive effective mass, \meff, computed by summing the signal leptons \pT, jets \pT\ and \met. It represents the total energy in the transverse plane, and is highly dependent on the initial particle mass. 
	\item The invariant mass of the electron and muon pairs that have different-charges ($m_{e^{\pm}e^{\mp}}, m_{\mu^{\pm}\mu^{\mp}}$, $m_\mathrm{SFOC}$). It helps to reject background events with $Z$ bosons.
	\item The sum of signal leptons \pT, $\sum \pT(\ell)$, and the sum of jets \pT, \sumjet, are variables that help given the high numbers of leptons and jets in the signal final state.
	\item The ratio of sum of the $b$-jets \pT, and the sum of all jets \pT, $\fracbjet$, is a very powerful discriminant when many $b$-jets are in the final state. 
	\item The minimum angular distance between the leading lepton and the selected jets, \DRlj. It is a key variable used to separate the UDD RPV signal illustrated in Figure~\ref{fig:diag_RPV1} from background events coming from e.g. \ttbar\ processes.
	\item The angular distance between the two same-charge leptons, $\Delta R (\ell^{\pm}, \ell^{\pm})$. It helps for the UDD RPV model, as the same-charge leptons are typically separated in the detector, for this signal process. 
\end{itemize}

\begin{table}[!tb]
	\centering
	\tbl{
		Exclusion and discovery signal regions defined for the bRPV model. $N_{\mathrm{B}}(\ell)$ ($N_{\mathrm{S}}(\ell)$) stands for the number of baseline (signal) leptons. Reused with permission from Ref.~\citen{ATLAS-CONF-2022-057}.
	}
		{\begin{tabular}{c|c|c}\toprule 
			& \SRtwolbrpv & \SRthreelbrpv \\
			\colrule 
			$N_{\mathrm{B}}(\ell)$ & \multicolumn{2}{c}{$\geq 2$} \\
			$N_{\mathrm{S}}(\ell)$ & = 2 & = 3 \\
			\pT($\ell$) & \multicolumn{2}{c}{$\geq 20$~GeV for (sub)leading leptons}  \\
			Charge($\ell$) & same-charge & -- \\
			\colrule
			\mttwo & $\geq 60$~GeV & $\geq80$~GeV \\
			\met & $\geq 100$~GeV & $\geq 120$~GeV \\
			\meff & -- & $\geq 350$~GeV \\
			\nbJ	 & = 0 & -- \\
			$\nJ$ $(\pT > 25)$~GeV & -- & $\geq 1$ \\
			$\nJ$ $(\pT > 40)$~GeV & $\geq 4$ & -- \\
			$m_{e^{\pm}e^{\mp}}, \; m_{\mu^{\pm}\mu^{\mp}}$ & -- & $ \notin [81,101]$~GeV \\
			\botrule
	\end{tabular}}
	\label{tab:SR_bRPV}
\end{table}

\begin{table}[!tb]
	\centering
	\tbl{
		Discovery signal regions defined for the UDD RPV model with two same-charge leptons in the final state. Reused with permission from Ref.~\citen{ATLAS-CONF-2022-057}.
	}
		{\begin{tabular}{c|c|c|c|c|c|c|c|c}
			\toprule
			& \multicolumn{2}{c|}{} & \multicolumn{3}{c|}{} & \multicolumn{3}{c}{} \\ 
			& \multicolumn{2}{c|}{\SRonebjrpv} & \multicolumn{3}{c|}{\SRtwobjrpv} & \multicolumn{3}{c}{\SRthreebjrpv} \\
			& L & M & L & M & H & L & M & H \\
			& \multicolumn{2}{c|}{} & \multicolumn{3}{c|}{} & \multicolumn{3}{c}{} \\
			\colrule	
			$N_{\mathrm{B}}(\ell)$ & \multicolumn{8}{c}{= 2} \\
			$N_{\mathrm{S}}(\ell)$ & \multicolumn{8}{c}{= 2} \\
			$\pT(\ell)$ & \multicolumn{8}{c}{$> 25$~GeV} \\
			Charge$(\ell)$ & \multicolumn{8}{c}{same-charge} \\
			\colrule 
			\nbJ & \multicolumn{2}{c|}{= 1} & \multicolumn{3}{c|}{= 2} &  \multicolumn{3}{c}{$\geq$ 3} \\ 
			$\sum \pT(\ell)$ & \multicolumn{2}{c|}{$\geq 100$~GeV} & \multicolumn{3}{c|}{--} & \multicolumn{3}{c}{--} \\
			\met & $\geq 100$~GeV & $\geq {50}$~GeV & \multicolumn{3}{c|}{$\geq {80}$~GeV} & \multicolumn{3}{c}{$\geq {20}$~GeV} \\
			\colrule 
			\nJ ($\pT > 25$~GeV) & $\leq 2$ & = 2 or = 3 & $\leq 3$ & =3 or = 4 & $\geq 5$ and $\leq 6$ & $\leq 3$ & $\leq 3$ & $\leq 6$ \\
			$\fracbjet$ & $\geq 0.7$ & $\geq 0.45$ & $\geq 0.9$ & $\geq 0.75$ & -- & $\geq 0.8$ & $\geq 0.8$ & $\geq 0.5$ \\
			\sumjet & $\geq {120}$~GeV & $\geq {400}$~GeV & $\geq {300}$~GeV & $\geq {420}$~GeV & $\geq {420}$~GeV & -- & -- & $\geq {350}$~GeV \\
			\DRlj & $\leq 1.2$ & $\leq 1.0$ & $\leq 1.0$ &  $\leq 1.0$ &  $\leq 1.0$ &  $\leq 1.5$ & -- & $\leq 1.0$ \\
			$\Delta R (\ell^{\pm}, \ell^{\pm})$ & $\geq 2.0$ & $\geq 2.5$ & $\geq 2.5$ & $\geq 2.5$ & $\geq 2.0$ & $\geq 2.0$ & -- & $\geq 2.0$ \\
			\botrule
		\end{tabular}}
	\label{tab:SR_UDDRPV}
\end{table}

Two non-overlapping signal regions are defined for the bRPV model, one with exactly two same-charge leptons (\SRtwolbrpv), and one with exactly three leptons (\SRthreelbrpv). These regions (Table~\ref{tab:SR_bRPV}) are used both as discovery and exclusion signal regions. 

For the UDD RPV model shown in Figure~\ref{fig:diag_RPV1}, three sets of non-overlapping discovery signal regions are selected, as shown in Table~\ref{tab:SR_UDDRPV}.
Each set has two or three overlapping signal regions that are defined with two same-charge lepton final states, and with exactly one, exactly two and at least three $b$-jets in the event. 
These overlapping regions were optimized using a signal generated with $\ninoone$ mass set to 180~GeV ($\SRrpv_{2\ell \mathrm{n}b}$ L), 200~GeV ($\SRrpv_{2\ell \mathrm{n}b}$ M) or 300~GeV ($\SRrpv_{2\ell \mathrm{n}b}$ H).

\begin{figure}[!th]
	\begin{center}
		\includegraphics[width=0.7\columnwidth]{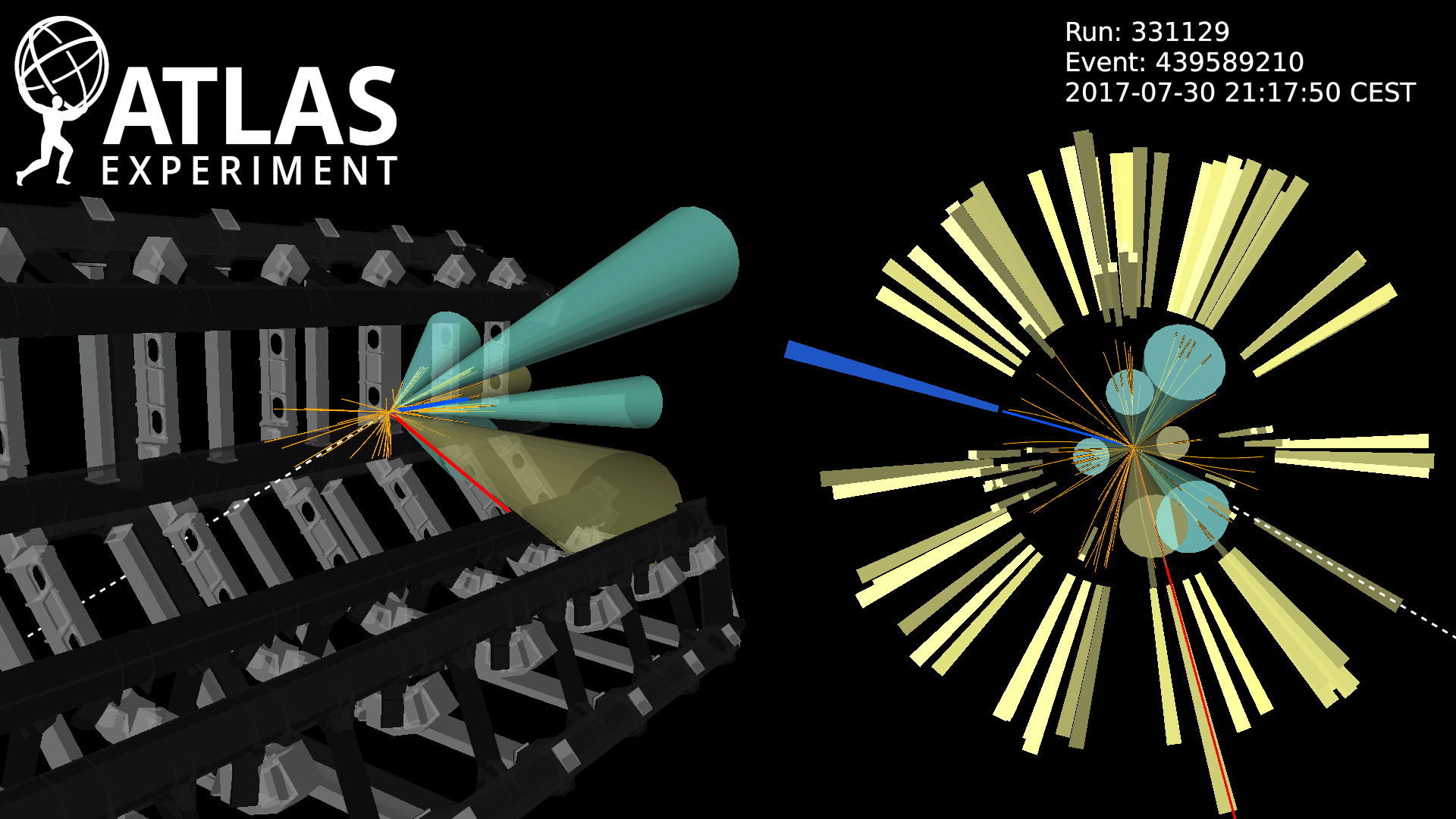}\vspace{-0.3cm}
	\end{center}
	\caption{
		Event display in a UDD RPV signal region, containing a muon and electron with same electric charge, and 6 jets. 
		The signal muon (electron) is indicated by the red (blue) line, and has $\pT = 35$~GeV (61~GeV). 
		The 6 jets from this event have a $\pT$ from 41~GeV to 145~GeV, and among them four are $b$-jets (shown with a cyan cone). 
		\met has a value of 31~GeV and is shown with a dotted white line.
		Figure reused with permission from Ref.~\citen{ATLAS-PPRPV1Lep}. 
	}
	\label{fig:evt_display}
\end{figure}

Selected exclusions signal regions for both UDD RPV production modes (Figures~\ref{fig:diag_RPV1} and~~\ref{fig:diag_RPV2}) are defined with two same-charge leptons, or with at least one lepton\footnote{The one lepton regions do not include events entering the two same-charge lepton regions.}.
In order to obtain the best sensitivity, events are further categorized into regions based on jet and $b$-jet multiplicities, and a neural network (NN) discriminant is introduced in some of the jet and $b$-jet multiplicity regions. 
These regions, and their background estimation, are discussed in detail in Ref.~\citen{ATLAS:2021fbt}.
For illustration, a (quite rare) data event display in a two same-charge signal region with six jets is shown in Figure~\ref{fig:evt_display}.

\begin{figure}[!tb]
	\begin{center}
		\includegraphics[width=0.73\columnwidth]{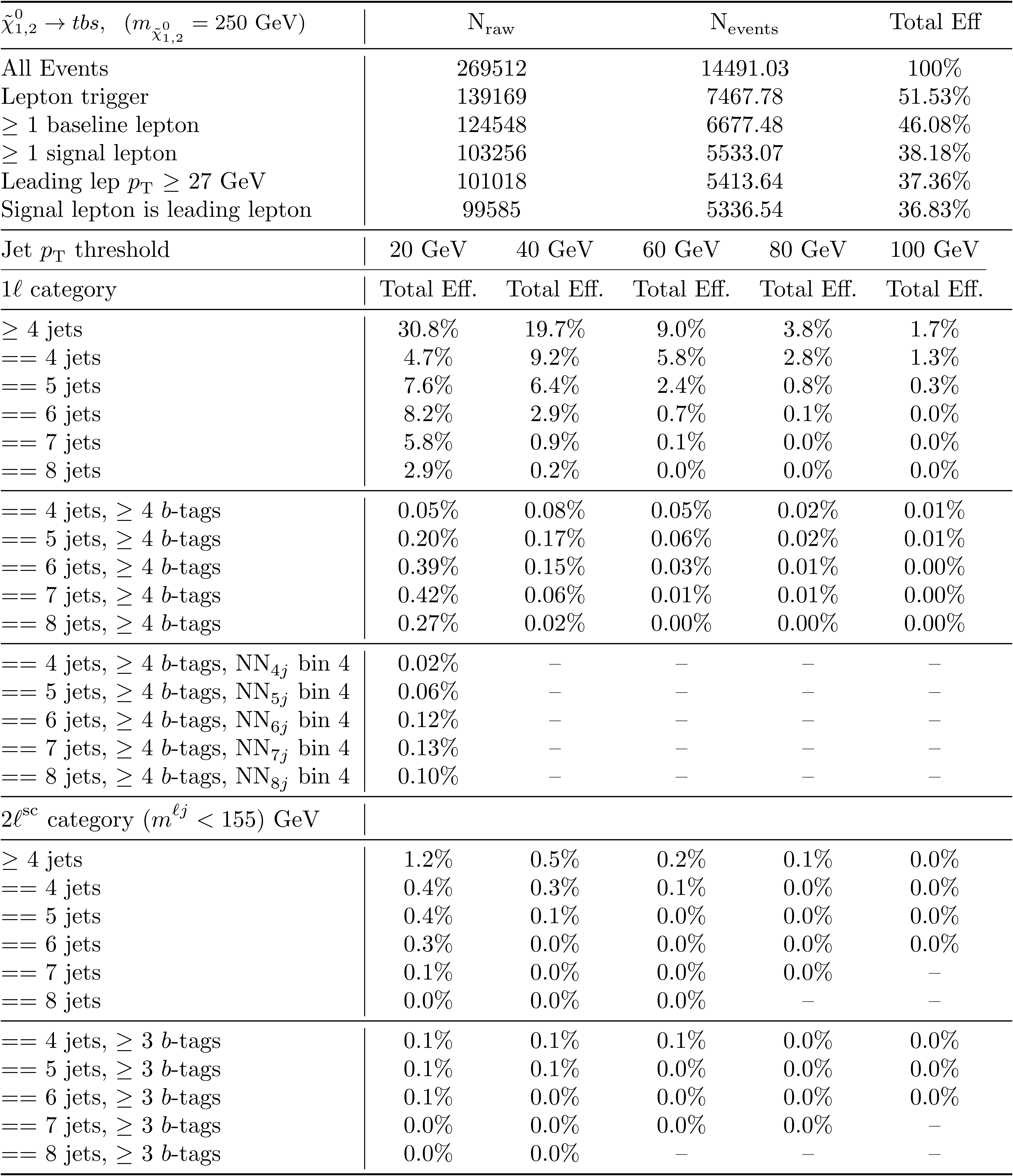}\vspace{-0.3cm}
	\end{center}
	\caption{
		Figure illustrating a cut flow table obtained for the RPV UDD model illustrated in Figure~\ref{fig:diag_RPV1}.
		$2\ell^{\mathrm{SC}}$ stands for two same-charge leptons, and 
		$m^{\ell j}$ is a variable that is very powerful at rejecting backgrounds that do not have lepton decays from two top quarks {(as the signal)~\cite{ATLAS:2021fbt}}.
		Selections that have not been evaluated in the analysis or are not applicable are denoted with a dash (–).
		Reused with permission from Ref.~\citen{ATLAS-PPRPV1Lep}.
	}
	\label{fig:cutflow}
\end{figure}

To illustrate the statistics obtained after key selections in the analysis, a cut flow is shown in the table from Figure~\ref{fig:cutflow}. 
It is obtained for the RPV UDD model illustrated in Figure~\ref{fig:diag_RPV1}, for a signal point generated with $\ninoone$ mass set to 250~GeV. 
The column labeled N$_{\mathrm{raw}}$ shows the number of generated events, while N$_{\mathrm{events}}$ shows the expected number of events with a luminosity of 139~fb$^{-1}$. 
The ``Total Eff" column shows the cut flow efficiency with respect to all weighted events.

\section{Backgrounds}
The background sources can be separated in two main categories.
The first category is populated by the SM processes that give one, two same-charge or three prompt lepton final states, like $W$+jets, $WZ$ or $\ttbar W$. 
In the second category are the detector backgrounds, such as events with electron charge flip or the fake/non-prompt leptons.
In this document, only the background strategy used to get the results in the signal regions shown in Tables~\ref{tab:SR_bRPV} and~\ref{tab:SR_UDDRPV} is discussed. The background studies done for the UDD RPV exclusion signal regions are presented in detail in Ref.~\citen{ATLAS-PPRPV1Lep}.

\subsection{$WZ$+jets background}
\begin{table}[!tb]
	\centering
	\tbl{
		The control and validation regions defined for the bRPV, and UDD RPV discovery signal regions. Reused with permission from Ref.~\citen{ATLAS-CONF-2022-057}.
	}
	{	\begin{tabular}{c|cccc}
			\toprule
			& \CRWZtwoj & \VRWZfourj & \VRWZfivej & \VRttV \\
			\colrule
			$N_{\mathrm{B}}(\ell)$ &  \multicolumn{3}{c}{= 3} & $\geq 2$ \\
			$N_{\mathrm{S}}(\ell)$ &  \multicolumn{3}{c}{= 3} & $\geq 2$ \\
			$p_T(\ell)$ &  \multicolumn{3}{c}{$\pT > {20}$~GeV for (sub)leading leptons} & $\pT > {30}$~GeV for same-charge leptons \\
			Charge$(\ell)$ &  \multicolumn{3}{c}{--} & same-charge \\
			\colrule
			\nbJ & = 0 & = 0 & = 0 & $\geq 1$ \\
			\nJ ($\pT\geq {25}$~GeV) & $\geq 2$ & $\geq 4$ & $\geq 5$ & $\geq 3$ with $\pT > {40}$~GeV \\
			\colrule
			\multirow{12}{*}{Additional requirements}
			& $50 < \met < {150}$~GeV & \multicolumn{2}{c}{$50 < \met < {250}$~GeV} & -- \\
			
			& $\meff < {1}$~TeV & \multicolumn{2}{c}{$\meff < {1.5}$~TeV} & -- \\
			
			& $81<m_\mathrm{SFOC}<101$~GeV & \multicolumn{2}{c}{$81<m_\mathrm{SFOC}<101$~GeV} & -- \\
			
			& \multicolumn{3}{c}{--} & $\Delta R (\ell_{1}, jet)_{\min}>1.1$ \\
			
			& \multicolumn{3}{c}{--} & $\sum\pT^{b\mathrm{-jet}}/\sum\pT^{\mathrm{jet}}>0.4$ \\
			
			& \multicolumn{3}{c}{--} & $\met/\meff > 0.1$ \\\\
			& \multicolumn{4}{c}{To ensure negligible signal contamination veto any events entering the bRPV signal regions, and require:} \\
			& \multicolumn{4}{c}{$\nbJ \geq 3$} \\
			& \multicolumn{4}{c}{$\nbJ \geq 1$, $\nJ \geq 4$ ($\pT>{50}$~GeV), $\met > {130}$~GeV} \\
			& \multicolumn{4}{c}{$\nbJ = 0$, $\nJ \geq 3$ ($\pT>{50}$~GeV), $\met > {130}$~GeV} \\
			& \multicolumn{4}{c}{$\nbJ = 0$, $\nJ \geq 5$ ($\pT>{50}$~GeV)} \\
			\colrule
			Purity & 85\% & 84\% & 77\% & 62\% \\
			\botrule
		\end{tabular}
	}
	\label{tab:CRVR}
\end{table}

The $WZ$+jets process is a dominant background in the 0~$b$-jet bRPV signal regions, and to correct a shape missmodeling seen in the jet {multiplicity~\cite{ATLAS:2019bsc}}, a dedicated control region is defined (Table~\ref{tab:CRVR}).
Using a background only fit, a normalization factor of $0.88\pm0.30$ is obtained, and further used to scale the $WZ$+jet MC simulations.
The accuracy of this approach is checked in two validation regions, \VRWZfourj and \VRWZfivej (Table~\ref{tab:CRVR}).
The other SM backgrounds are estimated using only MC simulations.
For the \ttZ\ and \ttW\ processes, one common validation region, \VRttV, is defined (Table~\ref{tab:CRVR}). 
This approach is motivated, as the signal regions themselves have a mixture of \ttZ\ and \ttW\ processes, as well as other top processes that contribute with prompt leptons.
Ideally, separate validation (and even control) regions should be defined for \ttW\ events, but is difficult given the large amount of fake/non-prompt leptons and the low statistics in data when applying selections closer to the signal regions.

\subsection{Electron charge flip background}
Backgrounds with electron charge flip, or with a wrong charge measurement, are relevant only for the same-charge lepton final states. 
When a high \pT\ electron interacts with the detector material, it can radiate a hard photon that converts into an electron-positron pair.
During the reconstruction, if the energy of the radiated electron is too small, then the energy deposited in the calorimeter can be matched with the radiated positron track.
If this happens, the charge of the initial electron is wrongly measured. 
Of course, this can be initiated also by positron candidates~\footnote{For simplicity, the generic term electron is further used both for electrons and positrons.}.
For muons, such processes are very rare, and the muon charge flip is found to be negligible.

\begin{figure}[!tb]
	\begin{center}
		\subfigure[]{\label{fig:ChFlip}
			\includegraphics[width=0.4\columnwidth]{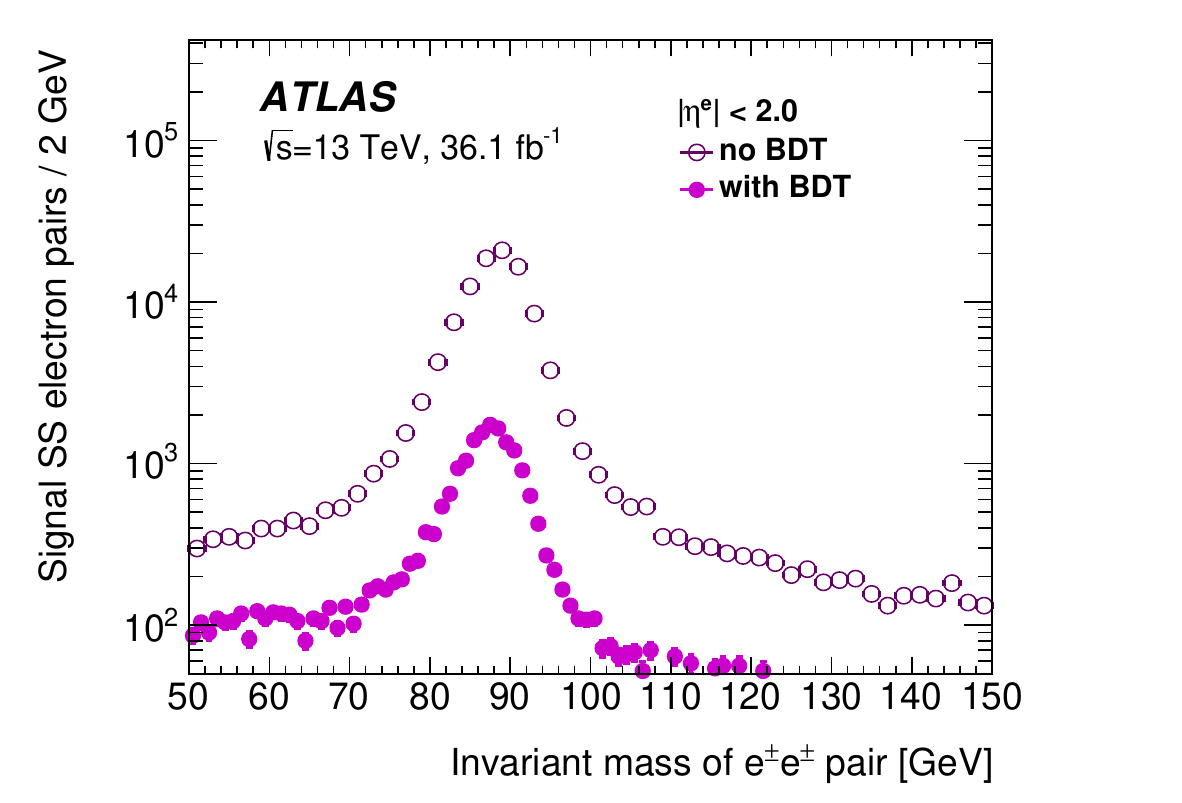}
		}\hspace{-0.7cm}
		\subfigure[]{\label{fig:Data_vs_Bkg}
			\includegraphics[width=0.6\columnwidth]{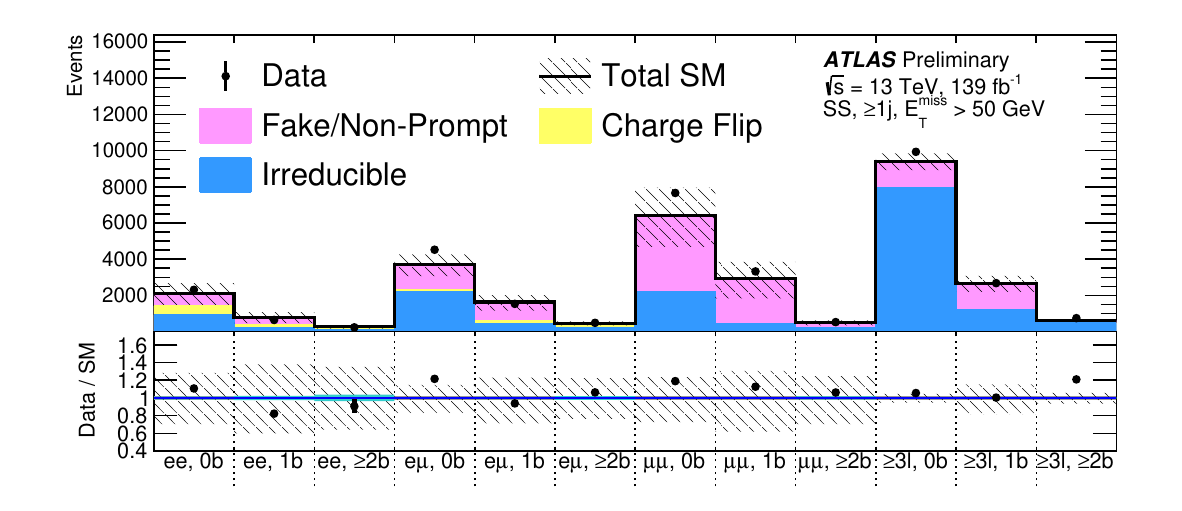}
		}\hspace{-0.7cm}\vspace{-0.4cm}
		\subfigure[]{\label{fig:Data_vs_Bkg_VRCR}
			\includegraphics[width=0.6\columnwidth]{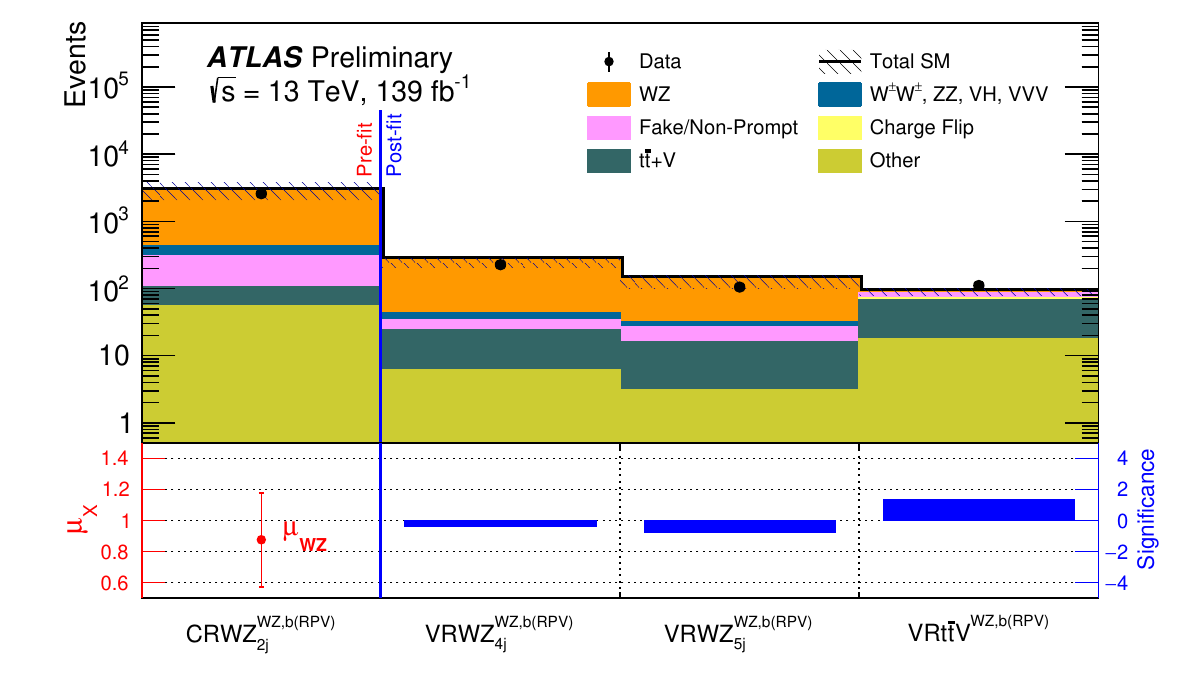}
		}
	\end{center}\vspace{-0.3cm}
	\caption{
		(a) Distribution of the invariant mass of signal $e^{\pm}e^{\pm}$ pairs with (full markers) and without (open markers) electron charge flip BDT selection applied. Reused with permission from Ref.~\citen{ATLAS_ChFlipBDT}.
		(b) Observed data compared with the background estimation after a loose preselection. The event yields are classified as a function of the lepton flavour and multiplicity, as well as the number of $b$-jets. The error bars include the statistical uncertainty and the full uncertainties for the data-based background estimates. Reused with permission from Ref.~\citen{ATLAS-CONF-2022-057}.
		(c) Results in control and validation regions.
		All uncertainties are included. Reused with permission from Ref.~\citen{ATLAS-CONF-2022-057}.\\
		SS stands for same-charge lepton pairs.
	}
	\label{fig:ChFlip_Data_vs_Bkg}
\end{figure}

Studies based on MC simulations show that in the signal regions, this type of background originates mainly from $Z\to e^{\pm}e^{\mp}$ (in $\SRtwolbrpv$) and dileptonic \ttbar\ processes.
As illustrated in Figure~\ref{fig:ChFlip}, this type of background is highly suppressed with a boosted decision tree discriminant (BDT) exploiting additional tracks in the vicinity of the electron and track-to-cluster matching {variables~\cite{ATLAS:2019qmc}}.
Thus, requirements on the BDT output are applied for both baseline and signal electrons.

To estimate the electron charge flip background in a given region, data events with different-charge leptons, but otherwise passing an identical selection, are taken and weighted with the charge flip probability, $\xi$. This weight can be written as $w_\mathrm{flip} = \xi_1(1-\xi_2) + (1-\xi_1)\xi_2$, with $\xi_i$ set to zero if the lepton $i$(=1 or 2) is a muon. 
$\xi$ is computed with a likelihood based {method~\cite{ATLAS:2019qmc}}, using $Z\to e^{\pm}e^{\mp}$ events, in bins of electron $\pT$ and $\eta$.
Checks are done using also \ttbar\ MC simulations.
The electron charge flip probabilities are found to be $\mathcal{O}(10^{-6})$ in the low $\pT$~-~$\eta$ regions, increasing to  $\mathcal{O}(1\%)$ in high $\pT$~-~$\eta$ regions.
Main sources of uncertainties on the $\xi$ parameter come from the low statistics in the measurement data sample, from the final-state radiation multijet production background estimation, and from the tight requirements on the di-electron invariant mass imposed to ensure a cleaner $Z\to e^{\pm}e^{\mp}$ data selection.
All the statistical and systematic uncertainties estimated for $\xi$ are propagated to the final electron charge flip estimation, leading to a 5\% to 45\% uncertainty in the predicted signal region yields for this background source.

\subsection{Fake/non-prompt lepton background}
The fake/non-prompt (not isolated) leptons stem from weak hadron decays, photon conversions in detector material or trident electrons.
Most common properties shared by these objects are an incorrect response to the lepton identification requirements, non-zero impact parameters, and are not well isolated.
These characteristics are used to discriminate the fake/non-prompt leptons against the prompt and isolated (real) leptons, and ultimately to estimate this source of background.
According to MC simulations, in the signal regions, this type of background originates mainly from  semi-leptonic or di-leptonic \ttbar\ processes.
The cases with a prompt leading lepton and a fake/non-prompt sub-leading lepton are dominant, and events with two fake/non-prompt leptons are negligible.

This type of background is evaluated with the data-based matrix {method~\cite{ATLAS:2022swp,ATLAS:2019fag}}, and cross-checked with a MC based method (MC-Template).
The matrix method relies on loose and tight lepton categories:
the loose (tight) leptons are baseline lepton candidates that fail (pass) the signal lepton requirements.
It relates the number of events containing prompt or fake/non-prompt leptons to the number of observed events with tight or loose leptons, using the probability for loose prompt or fake/non-prompt leptons to satisfy the tight lepton criteria.

The real lepton efficiency, or the probability for prompt leptons to pass the signal lepton requirements, is measured in \ttbar\ MC simulations using generator level information, as a function of \pT\ and $\eta$.
These ``true" efficiencies are corrected with dedicated scale factors that account for the data -- MC simulation differences in the reconstructed energy, or in the efficiency of identification and isolation of leptons.
For a $\pT$ around {15}~GeV they are found to be around 50--60\% (70\%) for electrons (muons), increasing up to 98\% (99\%) for leptons with $\pT>$~{100}~GeV ({60}~GeV).
Main uncertainties are from systematic sources associated to the correction scale factors, and come mainly from the lepton identification and isolation efficiency measurements.
The total uncertainties vary between 0.3--7\% (0.1--3\%) for electrons (muons), depending on $\pT$ and $\eta$.

The fake/non-prompt lepton efficiency, or the probability for fake/non-prompt leptons to pass the signal lepton requirements, is the most challenging to measure. 
The control regions enriched in this type of background should have a similar composition as the signal regions, similar kinematics, and be definable also in data. 
This is because not all sources of fake/non-prompt leptons are well modeled in MC simulations, thus an efficiency measurement in data is mandatory.
Six semileptonic or dileptonic \ttbar\ enriched control regions are defined with $e^\pm e^\pm$, $e^\pm \mu^\pm$, $\mu^\pm e^\pm$, $\mu^\pm \mu^\pm$, $\ell^\pm \ell'^\mp e^\mp$ and $\ell^\pm \ell'^\mp \mu^\mp$ lepton configurations, at least (one) two ($b$-) jets and $\met > 30$~GeV.
The minimum \met\ requirement helps to remove most of the QCD events with two fake/non-prompt leptons.
Upper cuts on \met\ and \meff\ are placed to minimize the signal contamination. 
The fake/non-prompt efficiency is measured per lepton flavor using the Tag~\&~Probe {method~\cite{ATLAS:2019qmc}}, in bins of \pT\ and $\eta$. 
The tag lepton must pass the signal lepton requirements and tighter identification and isolation operating points, to ensure it is the prompt lepton originating from one of the leptonically-decaying top quarks. 
In the measurement, the tag is taken to be the leading lepton from the same-charge pair.
The other lepton from the pair is the probe, and assumed to be the fake/non-prompt one.
The efficiencies in data are then simply obtained as the fraction of probe leptons satisfying the signal lepton requirements, after subtracting the expected contributions from SM processes with two or three prompt leptons and, when relevant, the electron charge flip.
They are measured independently in each control region, and a weighted combination gives the final efficiency.
The fake/non-prompt lepton efficiency has values of $\approx$10--20\% for both electrons and muons up to $\pT$ of around 45~GeV, and increases to 30--40\% for $\pT > {60}$~GeV. 
When two $b$-jets are present in the event, the fake/non-prompt lepton efficiencies are much higher, with an increase of up to a factor two.

\begin{table}[!tb]
	\centering
	\tbl{
		The systematic uncertainties associated to the fake/non-prompt lepton efficiency, measured in \ttbar\ MC simulation~\cite{ATLAS-CONF-2022-057,ATLAS:2019fag}.
	}
	{	\begin{tabular}{c|c|c}
			\toprule
			Source of uncertainty & Electrons & Muons \\
			\colrule
			Extrapolation to higher \pT &  \multicolumn{2}{|c}{$0\%$ (covered by measurement uncertainties and/or next item)}\\
			Underlying jet kinematics / event topology & $\pm30\%$ for $\pT<100$ GeV  & $\pm 30\%$ for $\pT<30$ GeV \\
			& $\pm50\%$ for $\pT>100$ GeV  & ${}^{+30\%}_{-50\%}$ for $30<\pT<50$ GeV \\
			&                              & ${}^{+30\%}_{-80\%}$ for $\pT>50$ GeV \\
			\botrule
	\end{tabular}}
	\label{tab:fakeNP_unc}
\end{table}

As shown in Table~\ref{tab:fakeNP_unc}, various sources of systematic uncertainties are considered, to account for all variations in composition and event {kinematics~\cite{ATLAS-CONF-2022-057,ATLAS:2019fag}}. 
The statistical uncertainty, the uncertainties associated to the electron charge flip subtraction, as well as a 30\% uncertainty on the SM background subtraction are also considered in the total uncertainty associated to the fake/non-prompt lepton efficiency. 
All these uncertainties are propagated through the matrix method to the fake/non-prompt background estimate, leading to a 10\% to 40\% uncertainty in the predicted signal regions yields for this background source.

The data and estimated background agree well, within the assigned uncertainties, as shown in Figure~\ref{fig:Data_vs_Bkg}.
The very good agreement in control and validation regions defined in Table~\ref{tab:CRVR} is illustrated in Figure~\ref{fig:Data_vs_Bkg_VRCR}: here
the electron charge flip background has a contribution only in the $ttZ/W$ validation region, as the $WZ$+jets regions are defined with three leptons.

The MC-Template method uses MC simulations to extrapolate the detector background predictions from control regions defined with low jet multiplicities and low \meff\ or \met, to the signal regions.  
The main assumption is that the MC simulations describe the kinematic distributions correctly and predict accurately e.g. the rate of fake/non-prompt leptons up to a global factor independent of the event kinematics and the process type.
This makes the method a suitable cross-check for the matrix method that assumes that the lepton fake/non-prompt efficiencies are the same in the control and signal regions, and independent of the selection requirements. 
The second assumption is that the fake/non-prompt fractions are uncorrelated in events with multiple fake leptons.

The MC-Template fake/non-prompt lepton control regions are similar to the ones used to measure the fake/non-prompt lepton efficiency needed by the matrix method.
As the $\ttW$ processes are not perfectly modeled at low \meff\ in MC simulations, a dedicated control region is defined.
The \CRWZtwoj\ region is also considered.
With these control regions six correction factors are measured, separately for $\ttW$, $WZ$, electron charge flip, fake electron, non-prompt electron, and fake/non-prompt muon contributions, via a simultaneous (1D or 2D) fit~\footnote{Note that the $\ttW$ correction factor obtained with the MC-Template method is applied to $\ttW$ MC simulations when the fake/non-prompt lepton efficiency is measured.}.
The fit uses a likelihood function defined as the product of the Poisson probabilities describing the observed events in the binned distributions from the expected number of events rescaled by the six correction factors which are left free to float in the fit.
Among the discriminant variables used are the leading and sub-leading lepton \pT and $\eta$, \meff or the jet multiplicity. 

Several sources of uncertainties are considered.
The statistical uncertainty is given by the fit, and corresponds to how much the correction factors need to be varied for one standard deviation change in the likelihood function. 
Systematic uncertainties are estimated by looking at the differences in the correction factors between various discriminant variables and/or control regions used in the fit.

\subsection{Background validation}
\begin{figure}[!tb]
	\begin{center}
		\subfigure[]{\hspace{-0.5cm}
			\includegraphics[width=0.256\columnwidth]{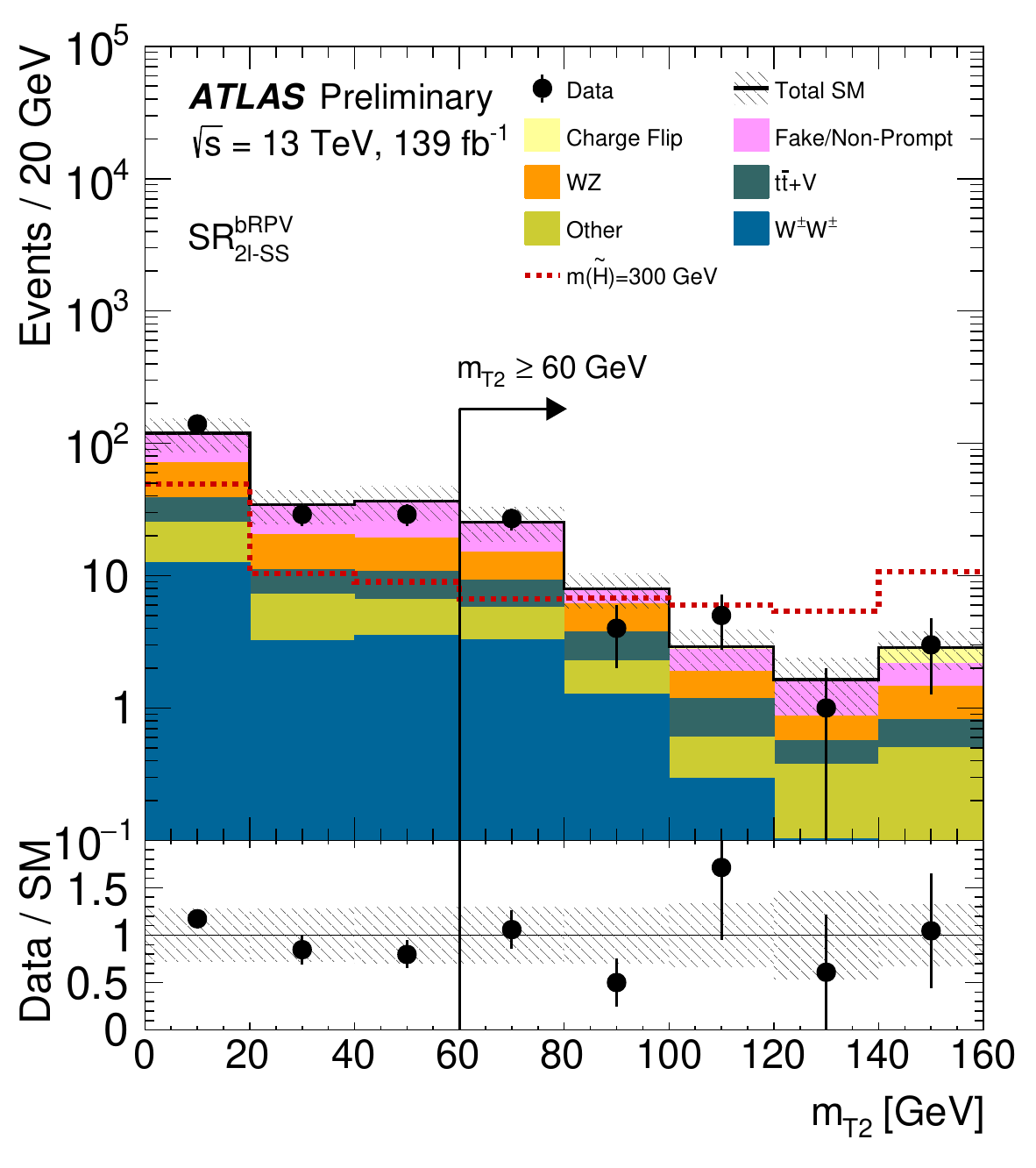}
		}\hspace{-0.45cm}
		\subfigure[]{
			\includegraphics[width=0.256\columnwidth]{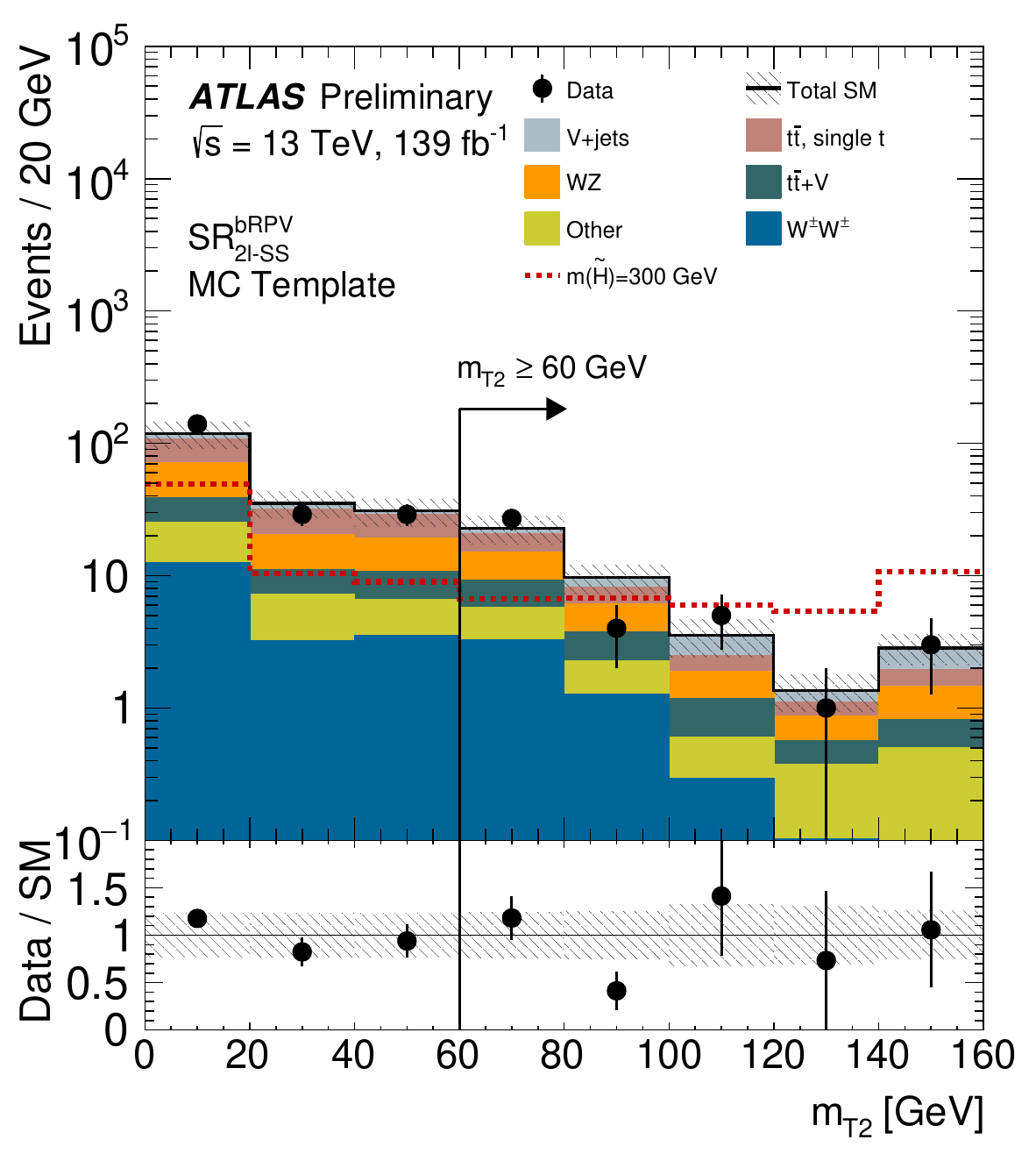}
		}\hspace{-0.45cm}
		\subfigure[]{
			\includegraphics[width=0.256\columnwidth]{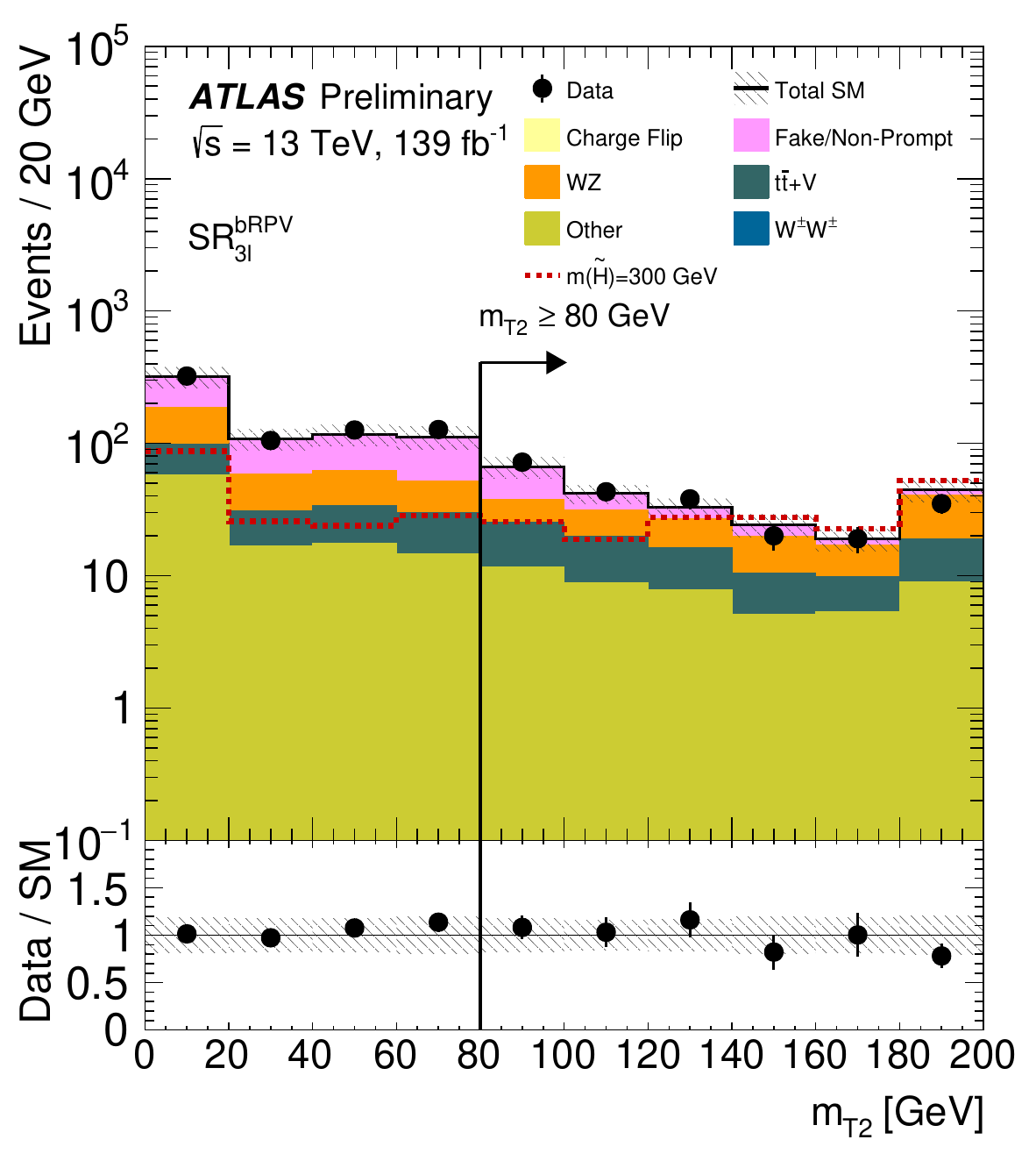}
		}\hspace{-0.45cm}
		\subfigure[]{
			\includegraphics[width=0.256\columnwidth]{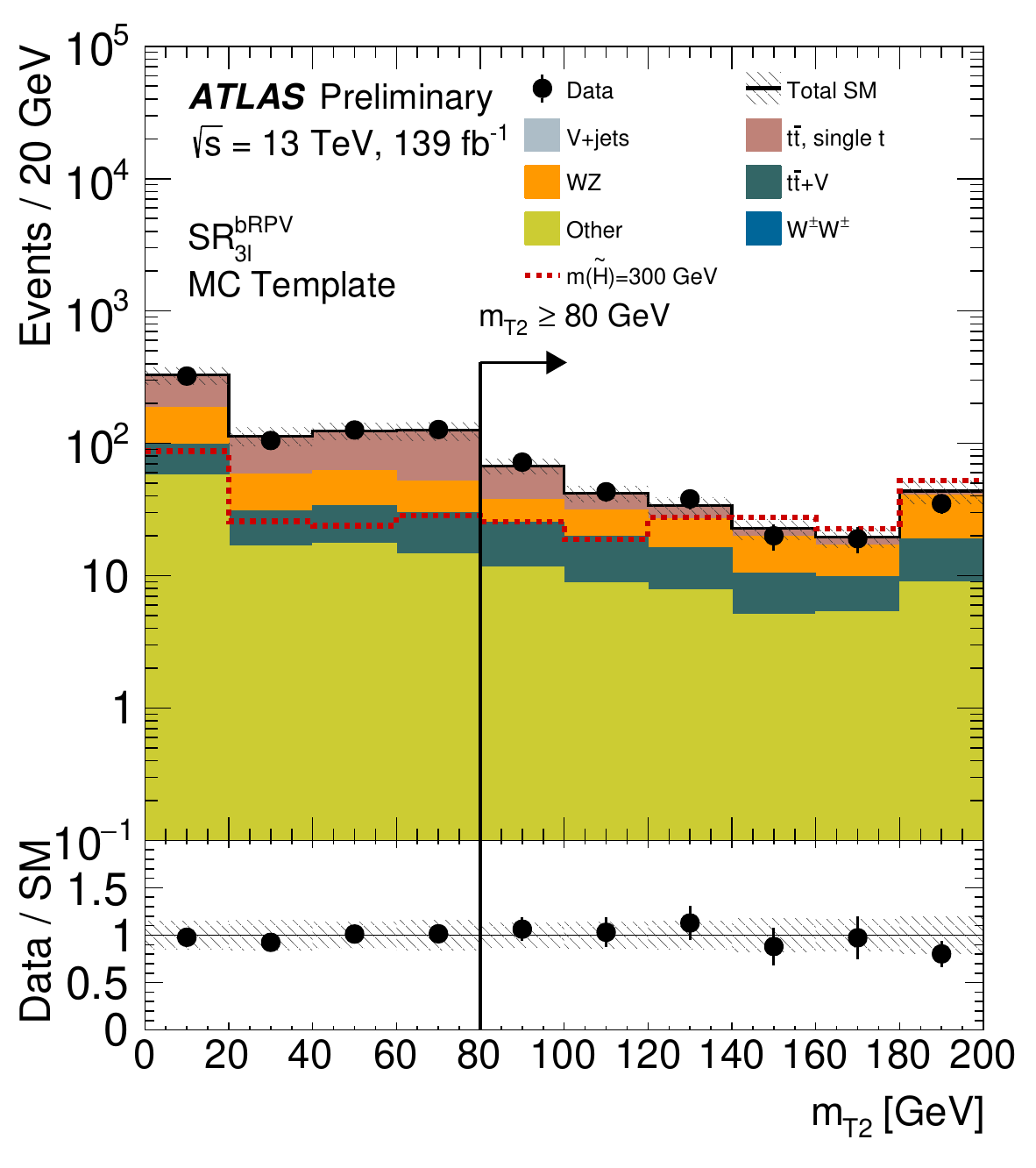}
		}
	\end{center}\vspace{-0.3cm}
	\caption{
		Observed data versus the estimated background close to the selected signal regions. 
		The bRPV signal regions requirements are all applied except for that on $\mttwo$. 
		Expected distributions of representative signal mass points are overlaid, and shown with interrupted red lines.
		In Figures a and c (b and d) the detector background is estimated with the data-based (MC-Template) methods described in the text.
		The vertical black lines and the corresponding arrows indicate the corresponding signal region.
		All uncertainties are included.
		In each figure the bottom panel shows the ratio of data to the estimated background.
		Reused with permission from Ref.~\citen{ATLAS-CONF-2022-057}.
	}
	\label{fig:bRPV_Nmin1}
\end{figure}
\begin{figure}[!tb]
	\begin{center}
		\subfigure{\hspace{-0.5cm}
			\includegraphics[width=0.256\columnwidth]{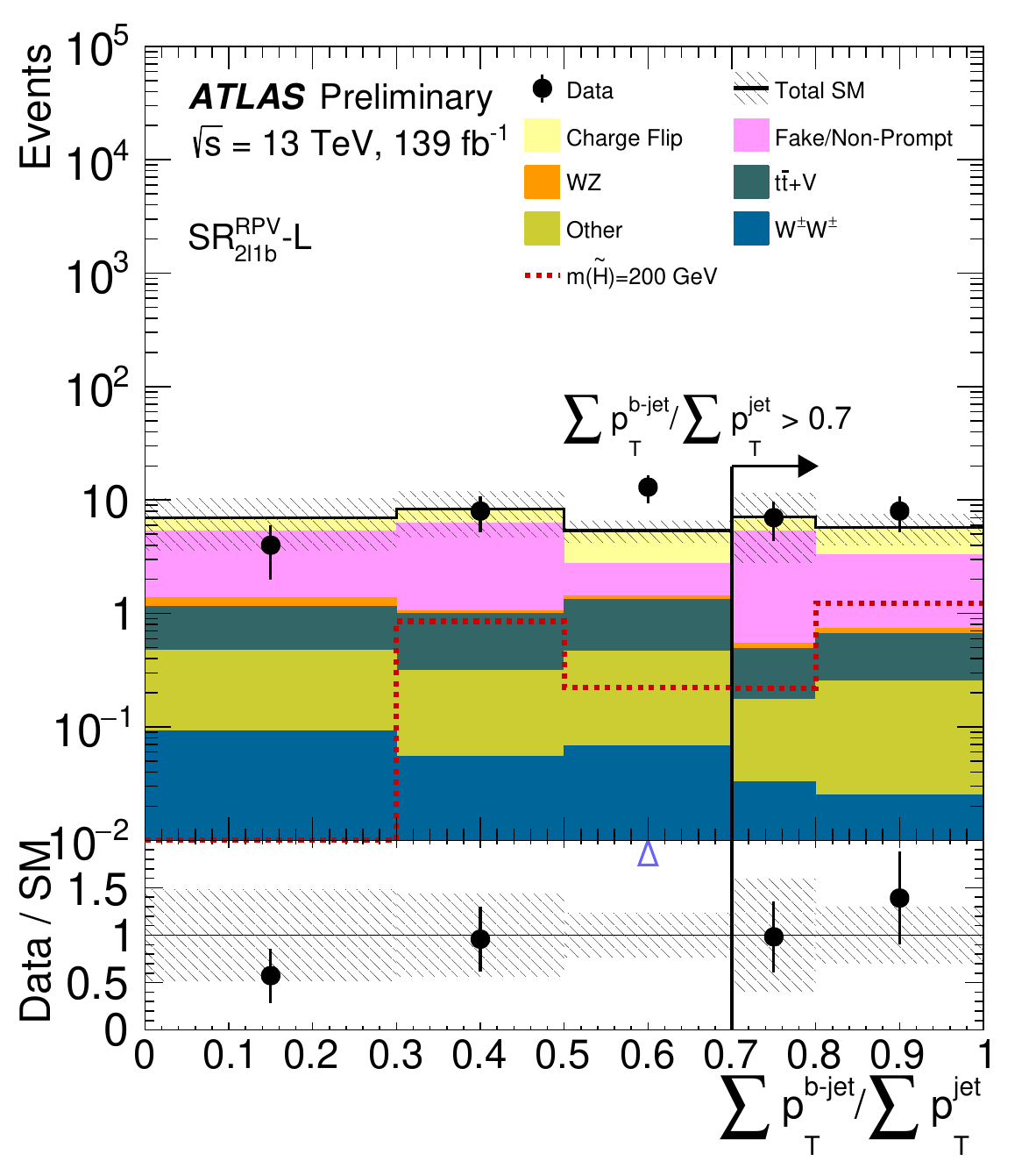}
		}\hspace{-0.45cm}
		\subfigure{
			\includegraphics[width=0.256\columnwidth]{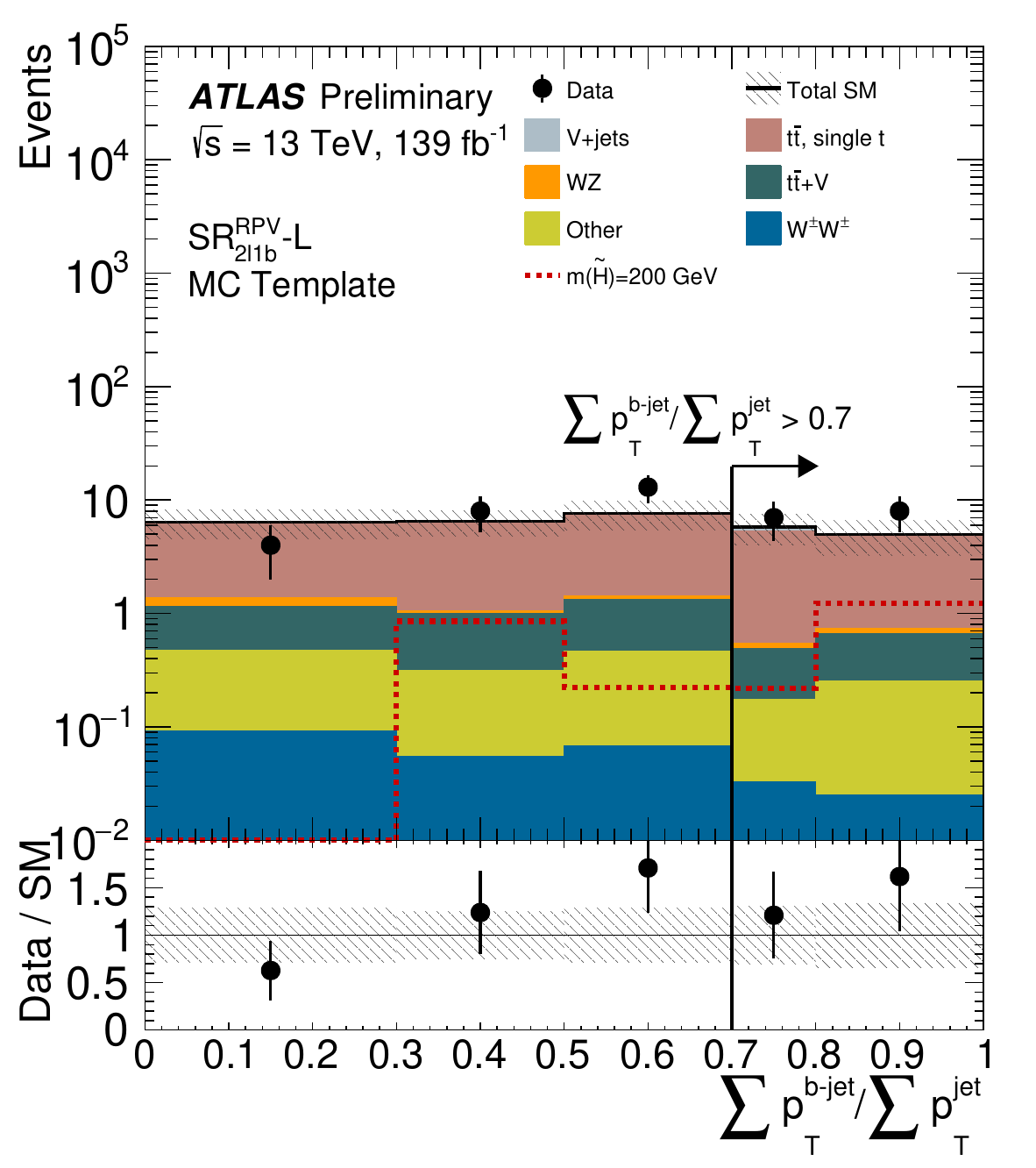}
		}\hspace{-0.45cm}
		\subfigure{
			\includegraphics[width=0.256\columnwidth]{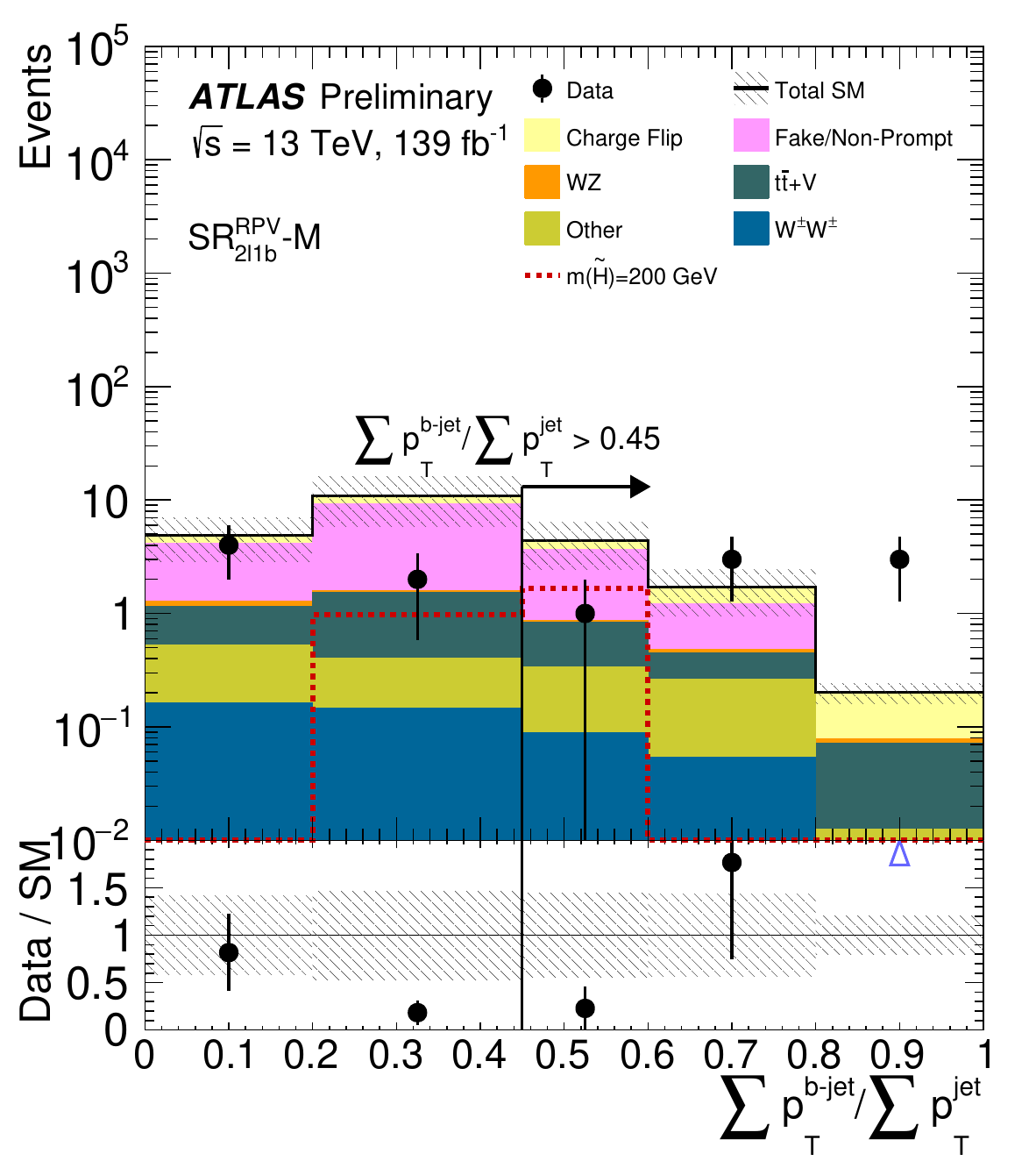}
		}\hspace{-0.45cm}
		\subfigure{
			\includegraphics[width=0.256\columnwidth]{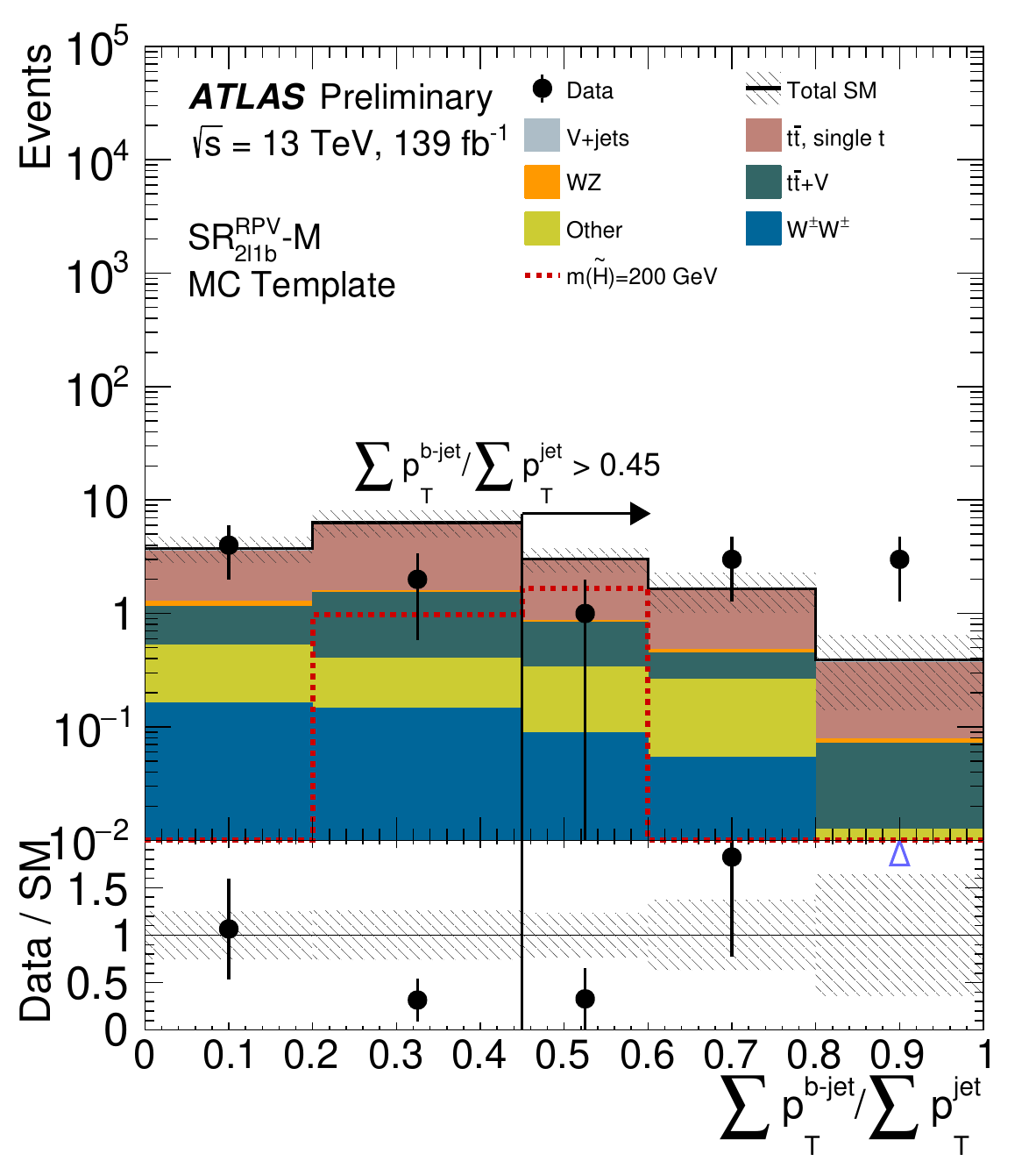}
		}
	\end{center}\vspace{-0.3cm}
	\caption{
		Similar to Figure~\ref{fig:bRPV_Nmin1}, but for the \SRonebjrpv signal {regions~\cite{ATLAS-CONF-2022-057}}. All the signal region requirements except the one on $\fracbjet$ are applied.
		Reused with permission from Ref.~\citen{ATLAS-CONF-2022-057}.
	}
	\label{fig:UDDRPV_Nmin1_set1}
\end{figure}
\begin{samepage}\nopagebreak
	\begin{figure}[!tb]
		\begin{center}
			\subfigure{
				\includegraphics[width=0.256\columnwidth]{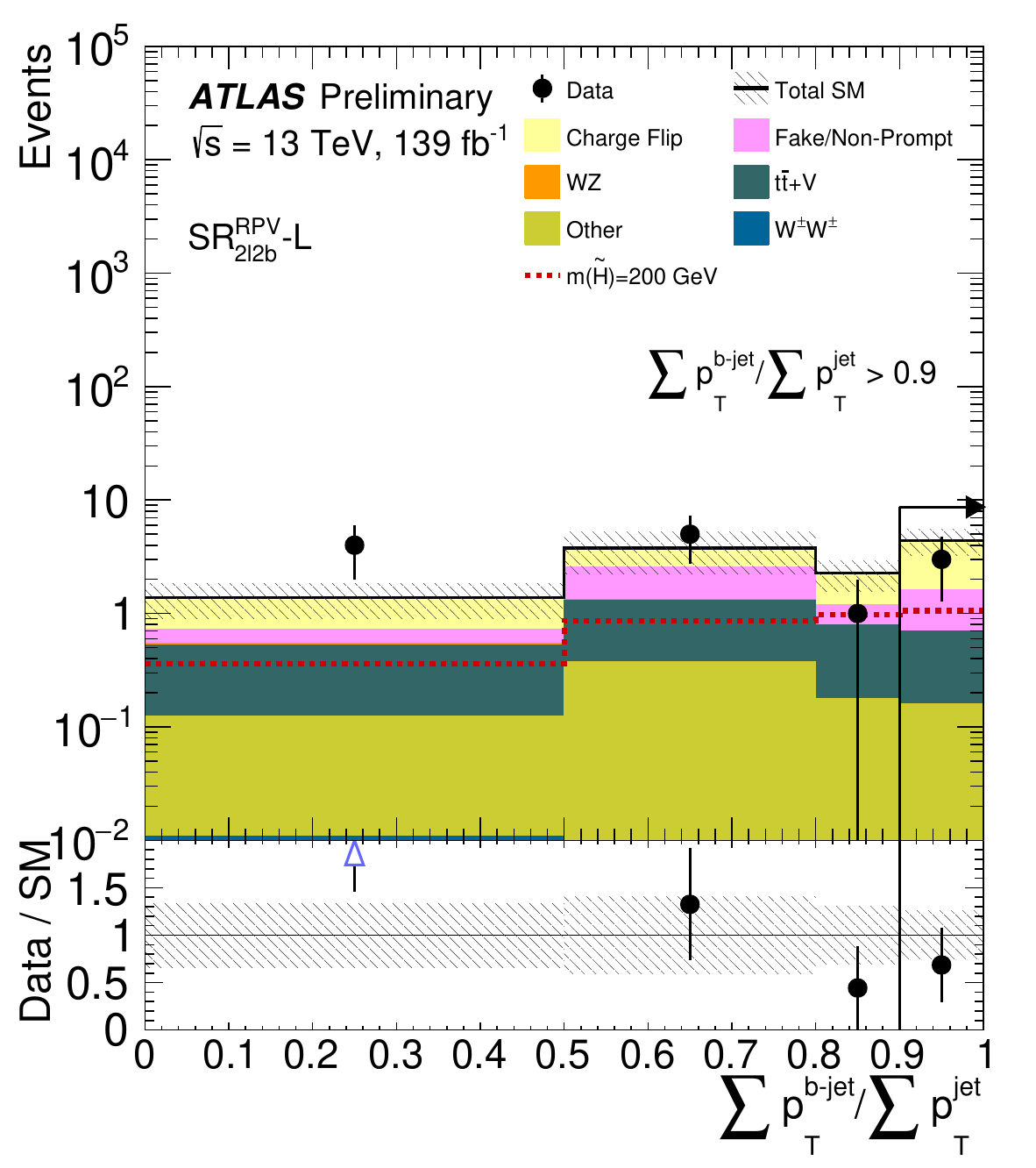}
			}\hspace{-0.45cm}
			\subfigure{
				\includegraphics[width=0.256\columnwidth]{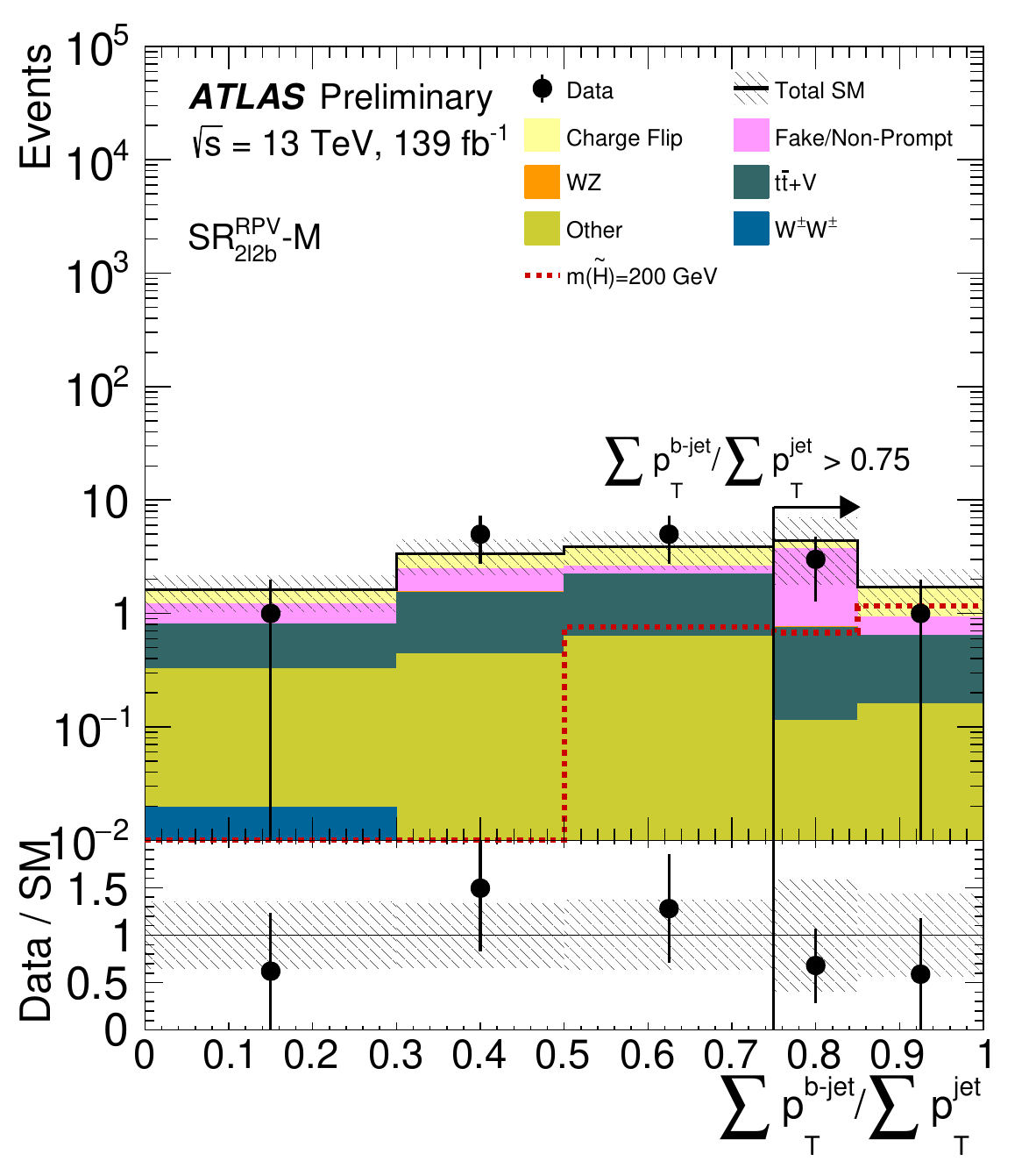}
			}\hspace{-0.45cm}
			\subfigure{
				\includegraphics[width=0.256\columnwidth]{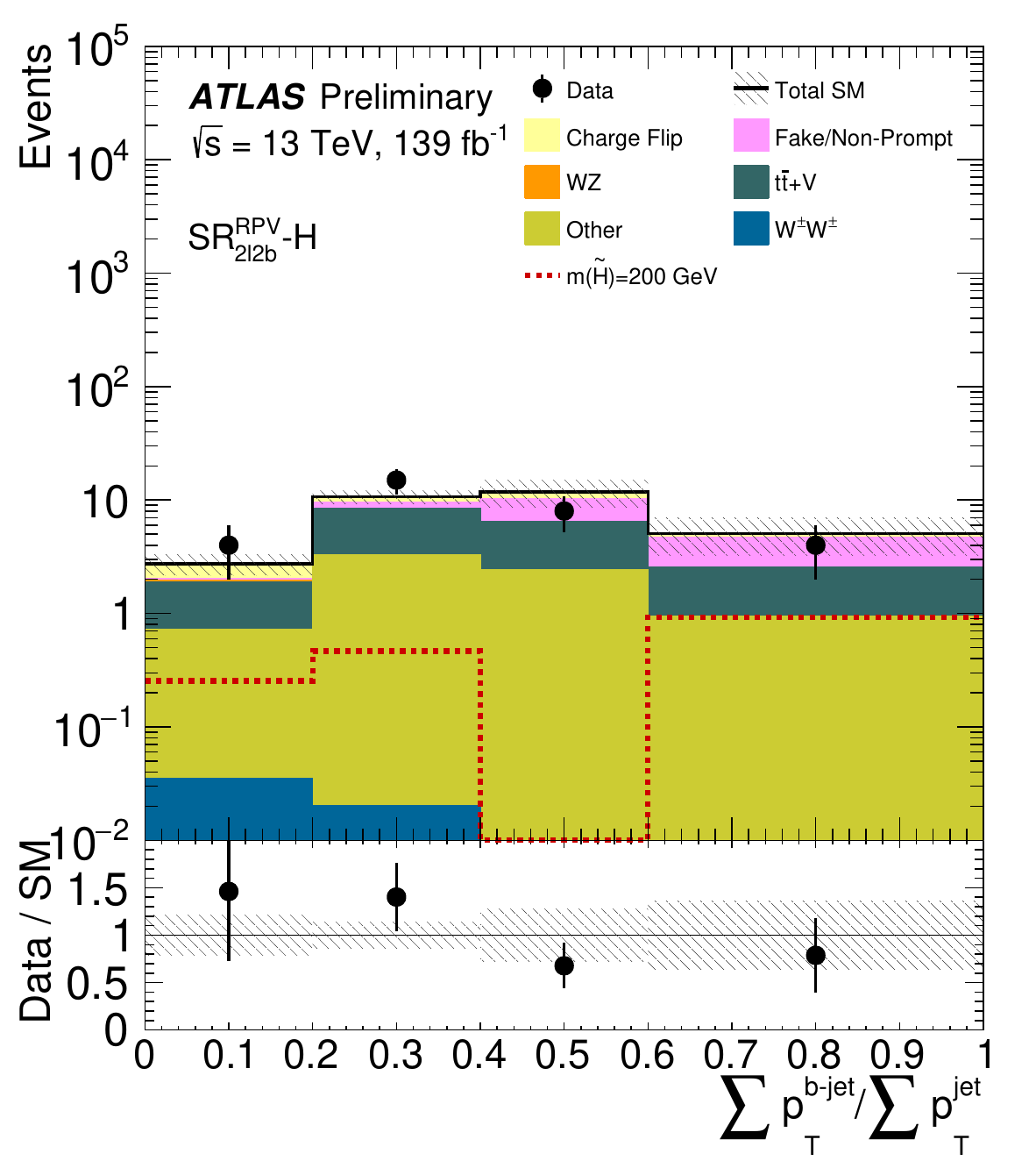}
			}\hspace{0.cm}\vspace{-0.4cm}
			\subfigure{
				\includegraphics[width=0.256\columnwidth]{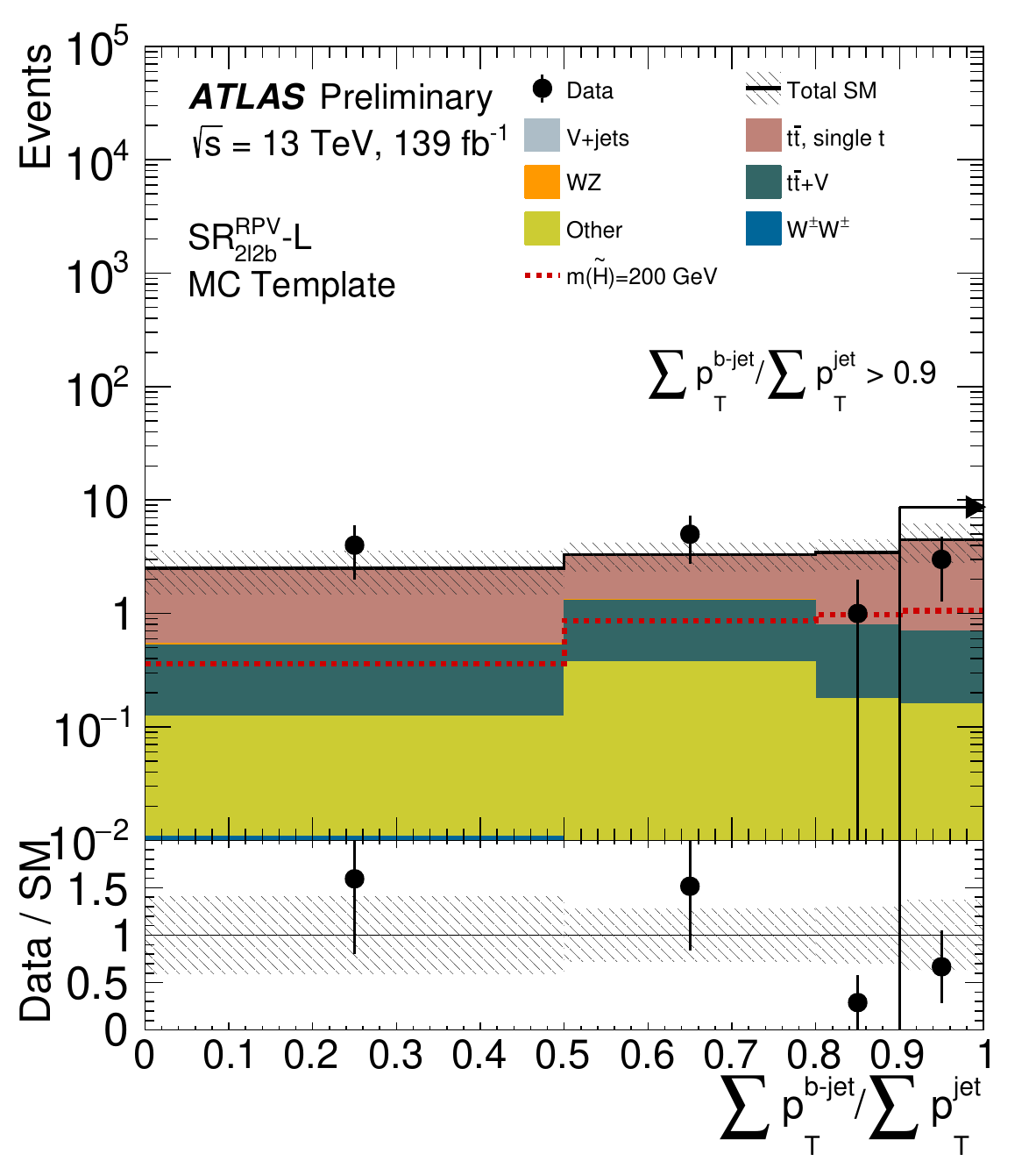}
			}\hspace{-0.45cm}
			\subfigure{
				\includegraphics[width=0.256\columnwidth]{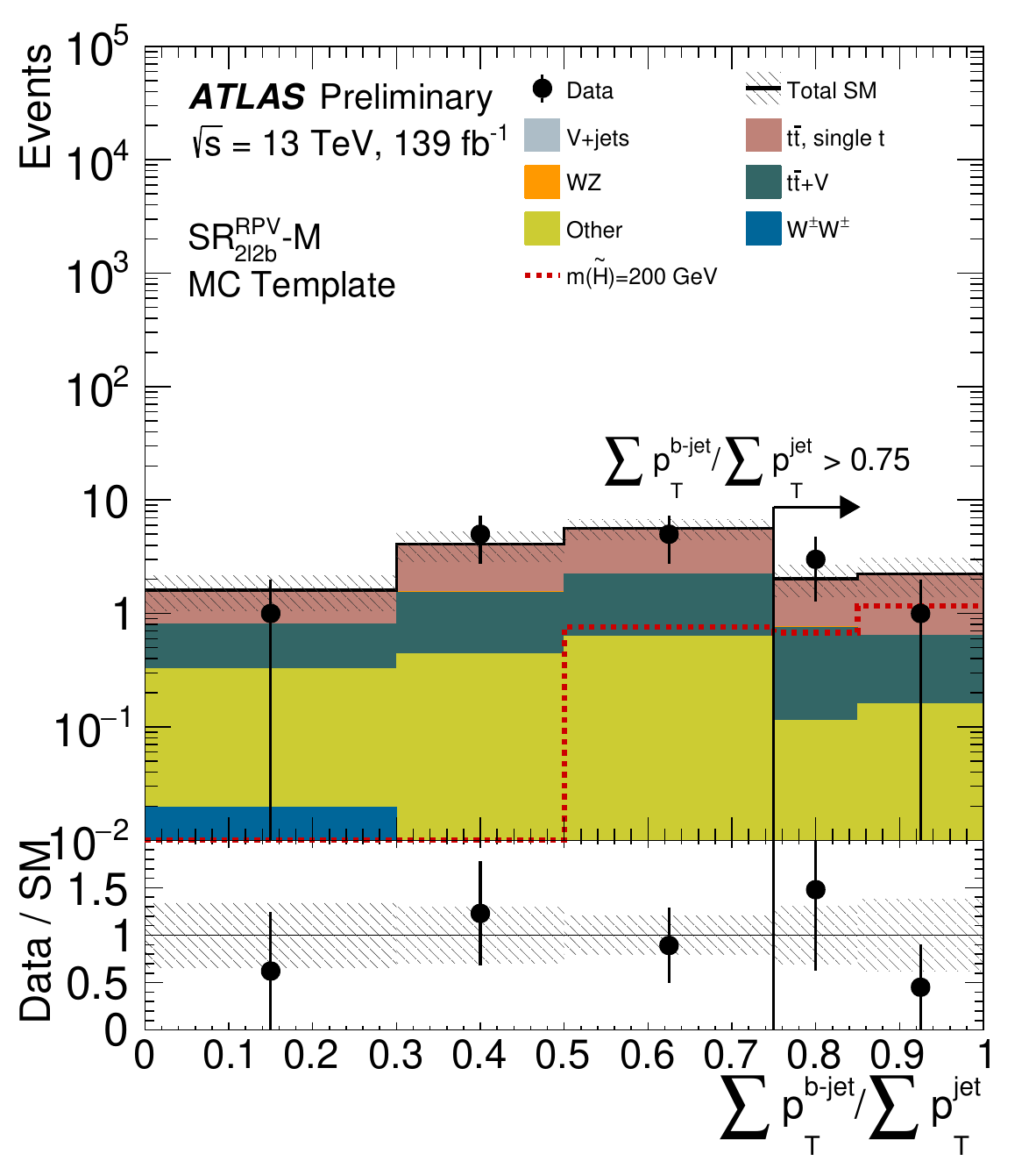}
			}\hspace{-0.45cm}
			\subfigure{
				\includegraphics[width=0.256\columnwidth]{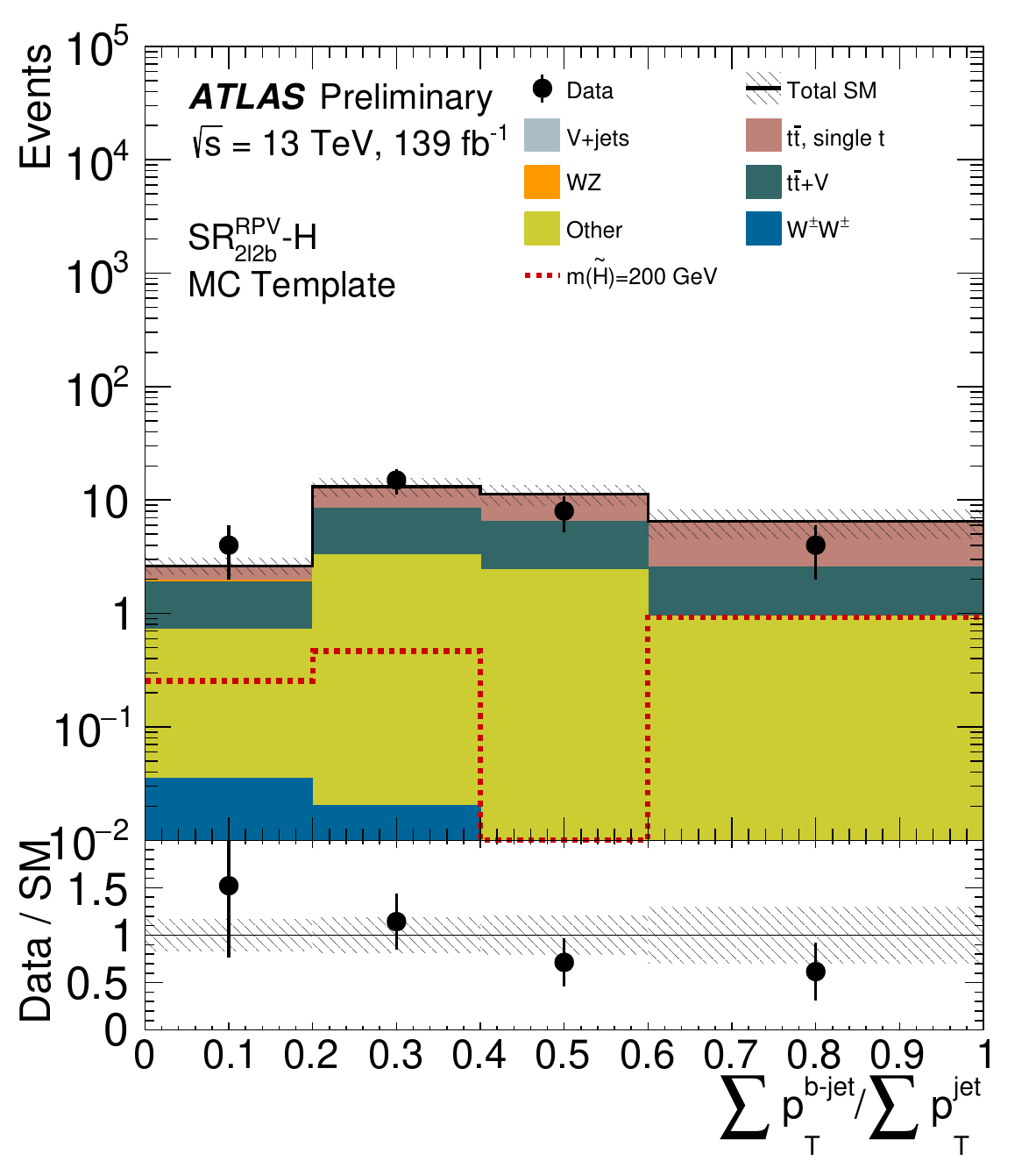}
			}
		\end{center}\vspace{-0.3cm}
		\caption{
			Similar to Figure~\ref{fig:UDDRPV_Nmin1_set1}, but for the \SRtwobjrpv signal regions. Top (bottom) the detector background is estimated with the data-based (MC-Template) methods described in the text.
			Reused with permission from Ref.~\citen{ATLAS-CONF-2022-057}.
		}
		\label{fig:UDDRPV_Nmin1_set2}
	\end{figure}
	\begin{figure}[!tb]
		\begin{center}
			\subfigure{
				\includegraphics[width=0.256\columnwidth]{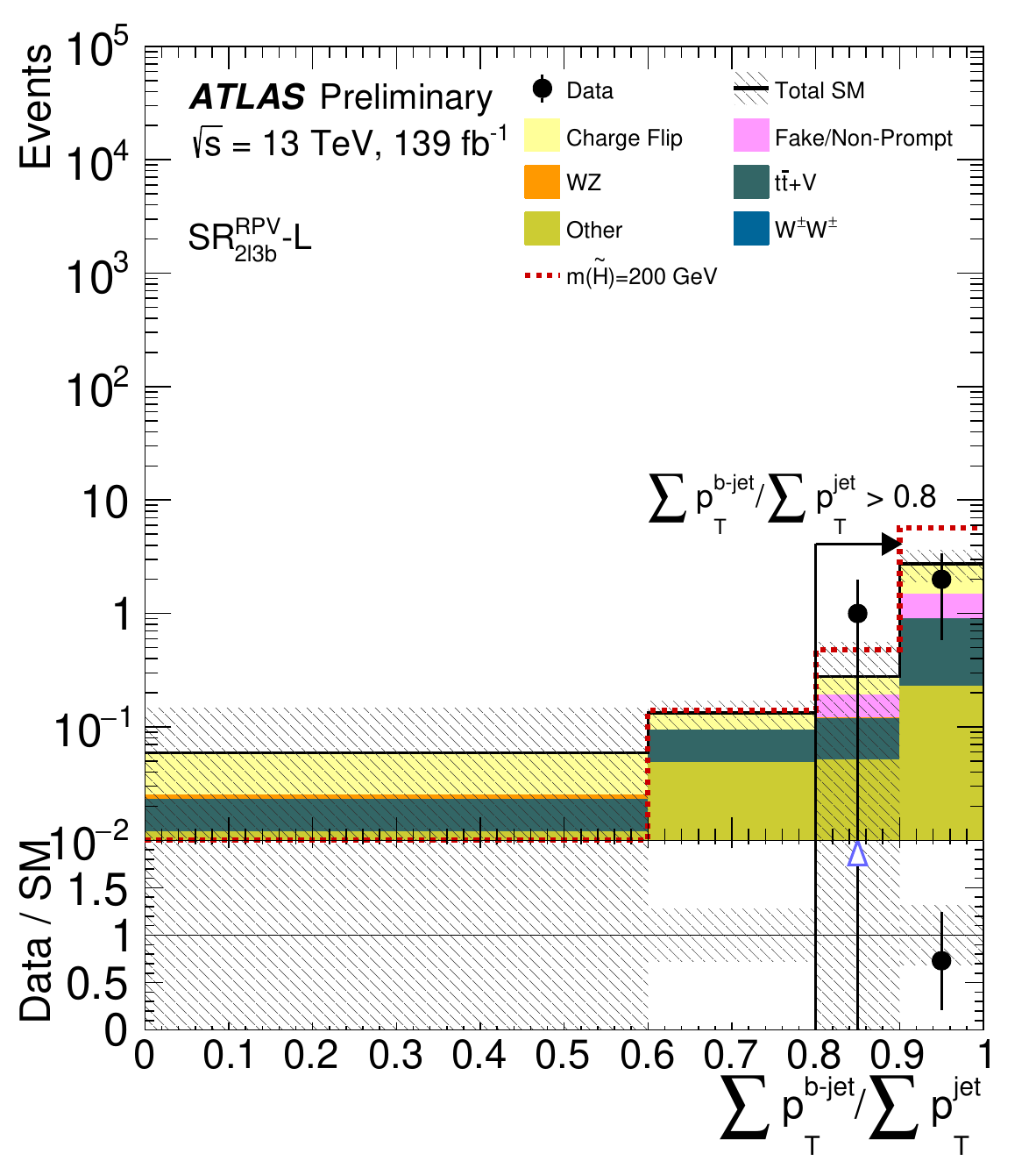}
			}\hspace{-0.45cm}
			\subfigure{
				\includegraphics[width=0.256\columnwidth]{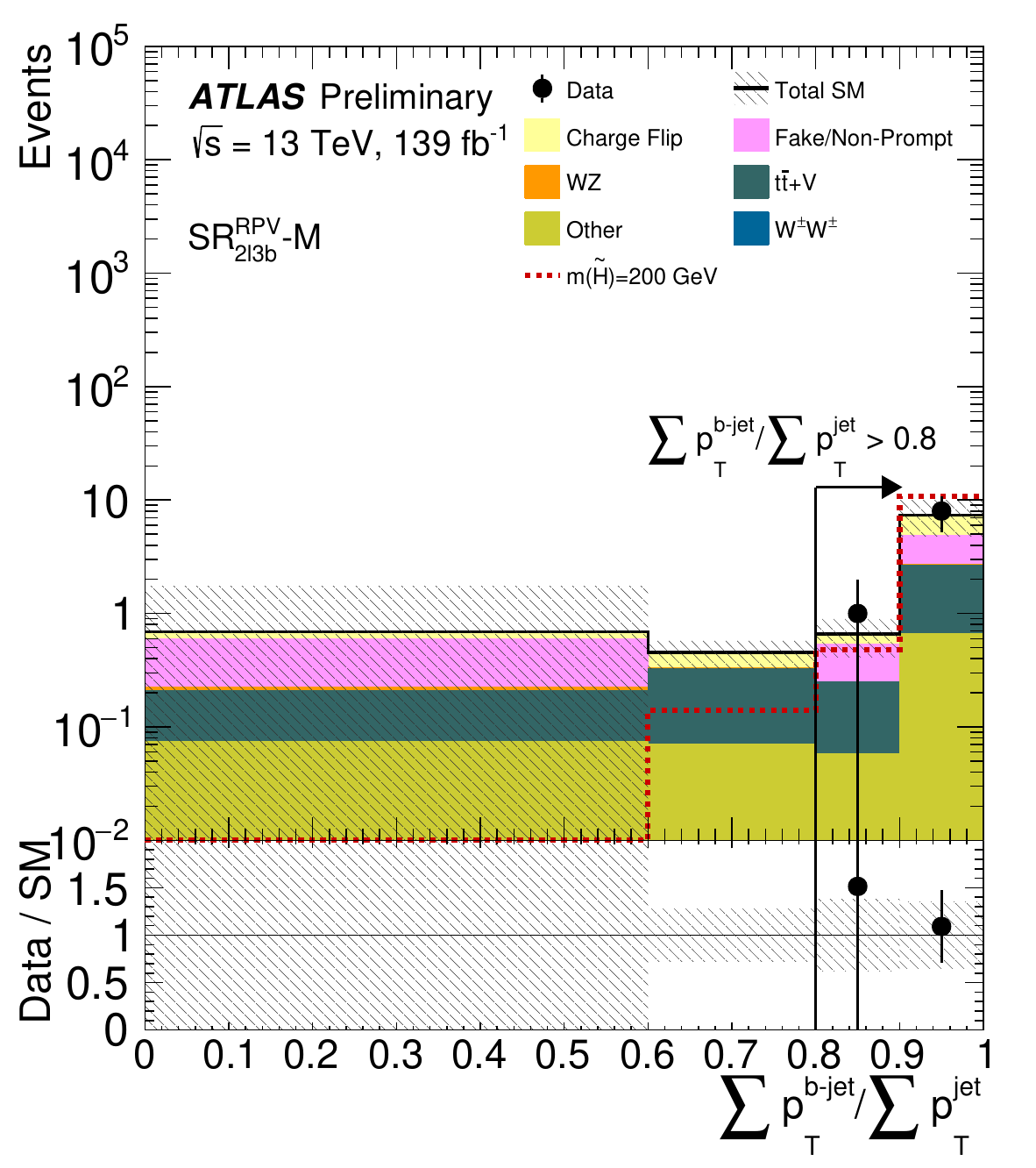}
			}\hspace{-0.45cm}
			\subfigure{
				\includegraphics[width=0.256\columnwidth]{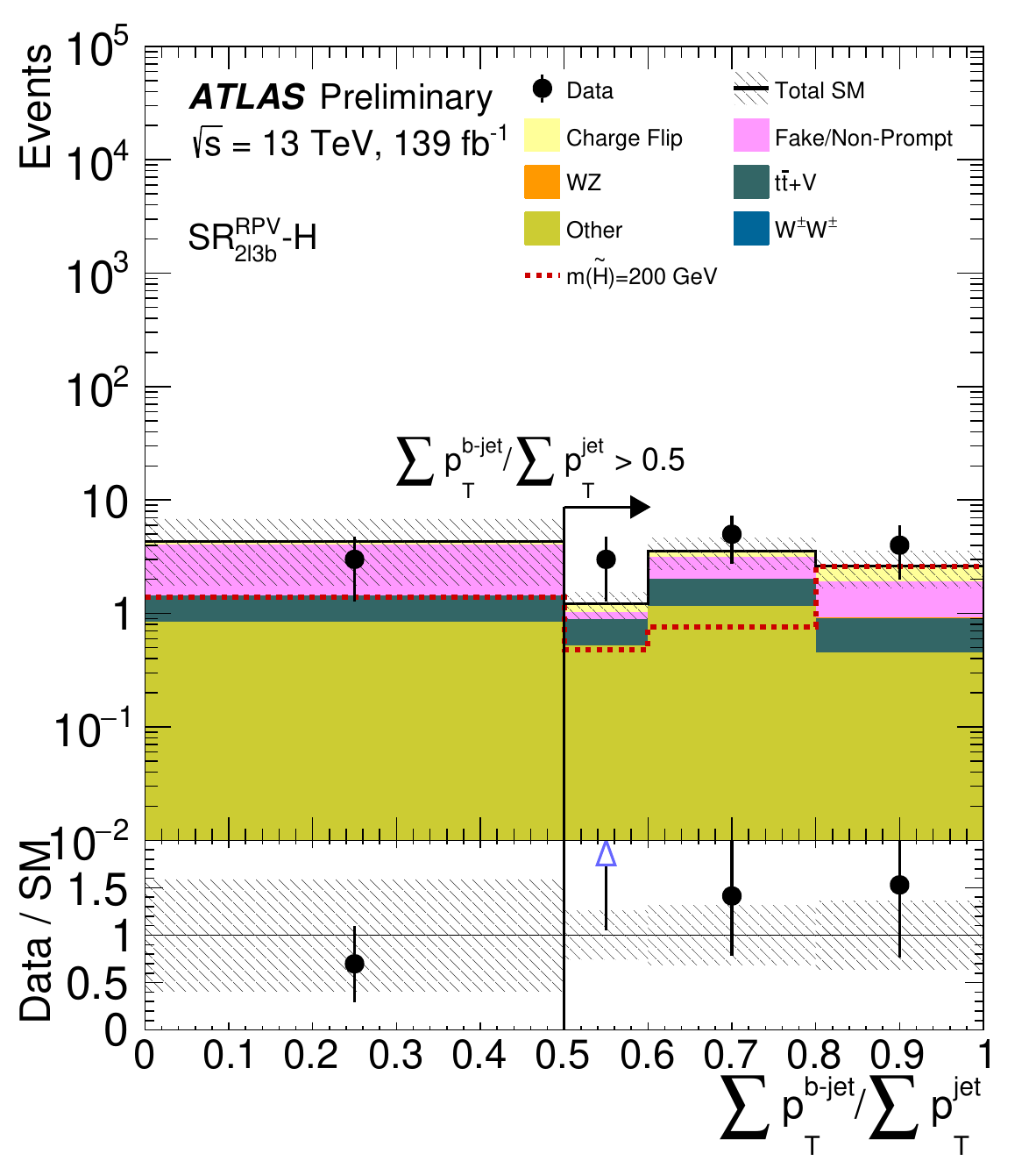}
			}\hspace{0.cm}\vspace{-0.4cm}
			\subfigure{
				\includegraphics[width=0.256\columnwidth]{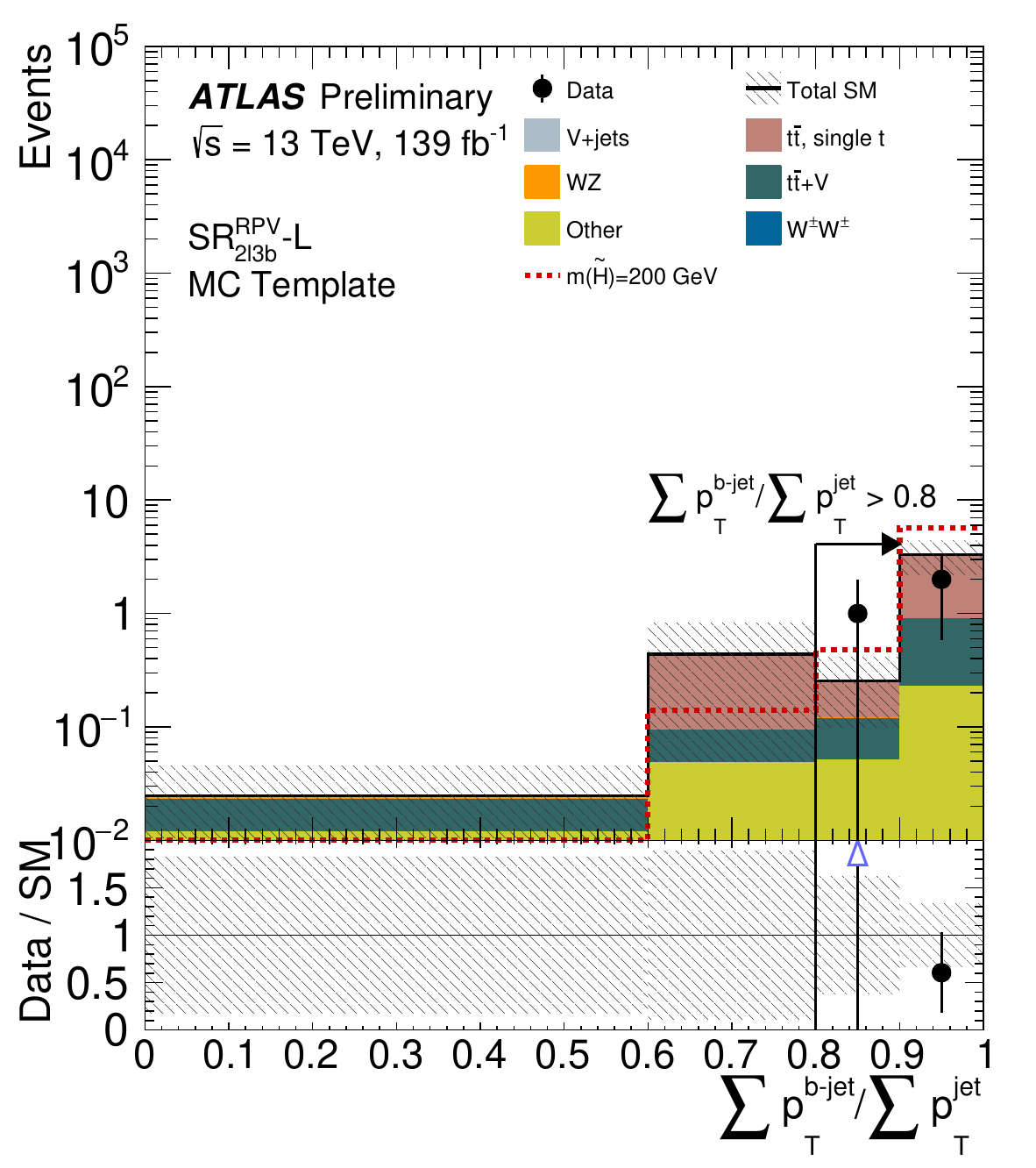}
			}\hspace{-0.45cm}
			\subfigure{
				\includegraphics[width=0.256\columnwidth]{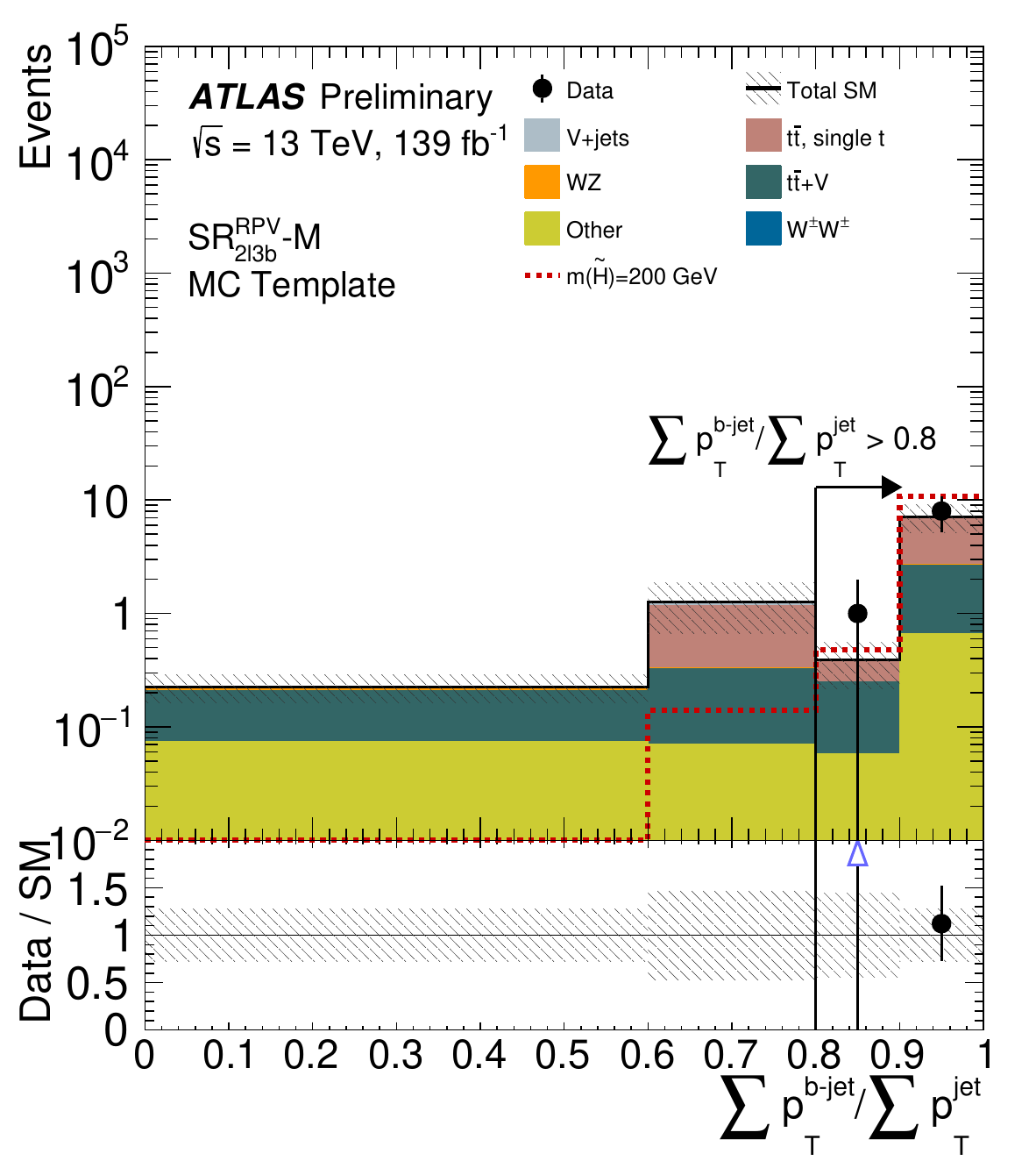}
			}\hspace{-0.45cm}
			\subfigure{
				\includegraphics[width=0.256\columnwidth]{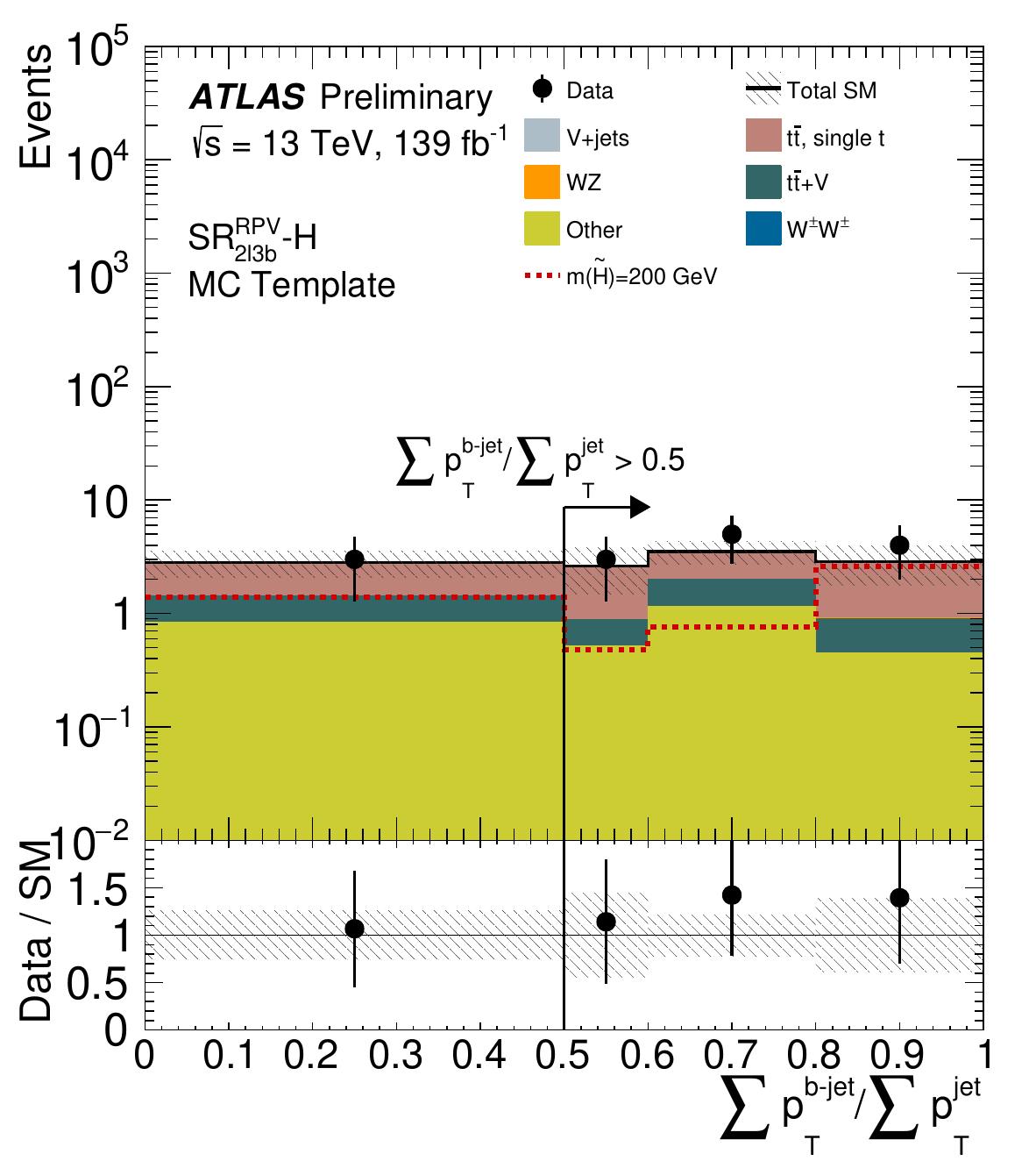}
			}
		\end{center}\vspace{-0.3cm}
		\caption{
			Similar to Figure~\ref{fig:UDDRPV_Nmin1_set2}, but for the \SRthreebjrpv signal regions. Reused with permission from Ref.~\citen{ATLAS-CONF-2022-057}. 
		}
		\label{fig:UDDRPV_Nmin1_set3}
	\end{figure}
\end{samepage}

The good observed data -- estimated background agreement close to the selected signal regions is illustrated in Figures~\ref{fig:bRPV_Nmin1}--\ref{fig:UDDRPV_Nmin1_set3}.
Here one can see also the good agreement between the detector background estimations with the data-based and MC-Template methods, 
proving that all methods used for the background estimation are robust.
Note also the results in the signal regions, highlighted with an arrow in the figures. 
In some regions, the MC-Template estimations have lower uncertainties and this will be studied in more detail in future.
One improvement will be to combine the fake/non-prompt lepton background estimations obtained with the matrix and MC-Template methods.
This will help not only with the reduction of uncertainties, but also with an improved estimation especially in the two same-charge leptons plus 2 or 3 $b$-jets regions.
The latter is hinted at by the data to background ratio seen in Figures~\ref{fig:UDDRPV_Nmin1_set2} and~\ref{fig:UDDRPV_Nmin1_set3}.

\section{Systematic uncertainties}
\begin{figure}[!tb]
	\begin{center}
		\subfigure[]{\label{fig:Unc_CRVRs}\hspace{-0.3cm}
			\includegraphics[width=0.5\columnwidth]{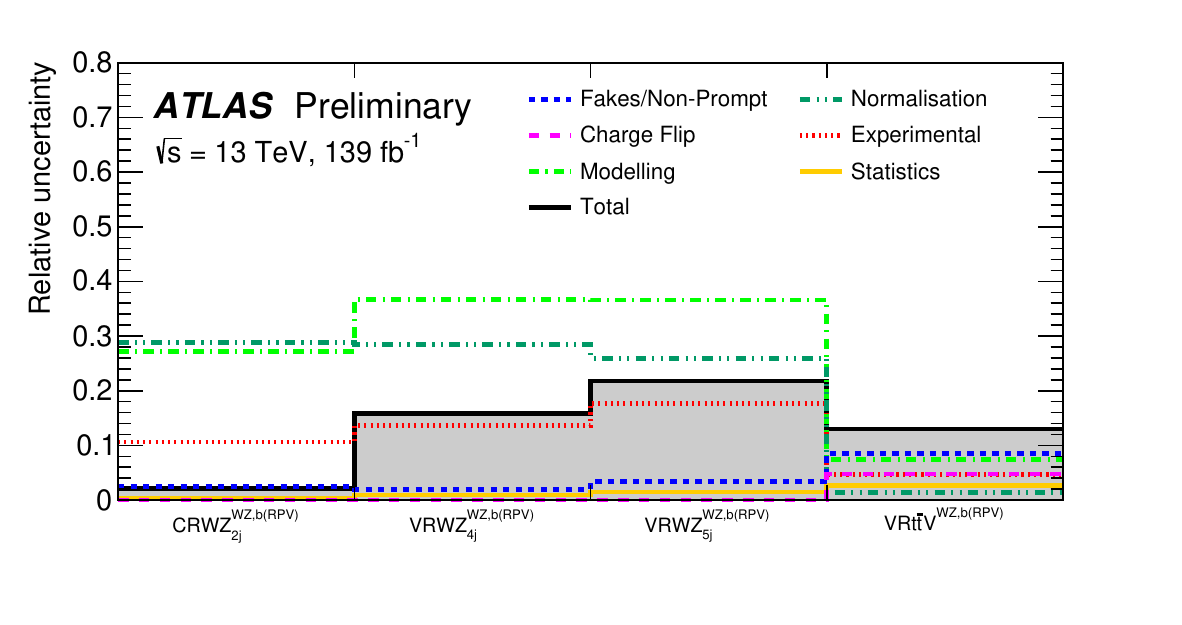}
		}\hspace{-0.45cm}
		\subfigure[]{
			\includegraphics[width=0.5\columnwidth]{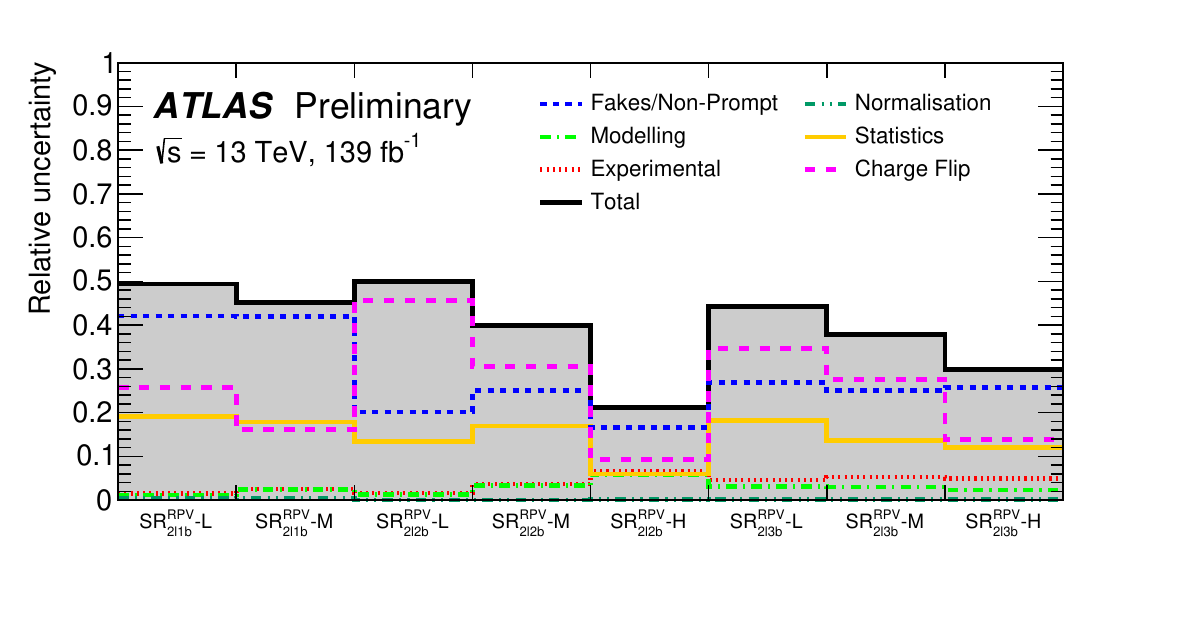}
		}
	\end{center}\vspace{-0.3cm}
	\caption{
		Contributions of different categories of uncertainties relative to the expected background yields in the (a) control and validation regions, and (b) in the UDD RPV discovery signal regions. 
		The total uncertainty takes into account also the correlation between different nuisance parameters.
		Reused with permission from Ref.~\citen{ATLAS-CONF-2022-057}.
	}
	\label{fig:Unc}
\end{figure}

Beside the uncertainties on the detector background estimates, all the experimental and theoretical sources of uncertainties are considered as detailed in Ref.~\citen{ATLAS-CONF-2022-057,ATLAS:2021fbt}. 
The breakdown of various uncertainty sources in the background prediction in the $WZ$+jets control region, in the $WZ$+jets and $ttZ/W$ validation regions, and in the UDD RPV discovery signal regions is shown in Figure~\ref{fig:Unc}.
In the ``Fakes/non-prompt" (``Charge-flip") category, all the systematic uncertainties associated to the fake/non-prompt lepton (electron charge flip) background are considered. 
In the ``Normalisation" category, all the systematic uncertainties associated to $WZ$+jets normalisation are considered, and in the ``Modelling" the systematic uncertainties associated to theoretical modelling for all SM backgrounds are also taken into account, respectively. 
As the name suggests, the ``Statistics" (``Experimental") category has all the statistical (experimental) uncertainties.

In the $WZ$+jets validation regions, the ``Normalisation" and ``Modelling" sources have a significant contribution (Figure~\ref{fig:Unc_CRVRs}). 
This is expected, as there is no fit in the validation regions, and the anti-correlations between the theory uncertainties are not accounted for. 
The analysis sensitivity is not affected too much, as most of the discussed discovery signal regions do not have as a dominant component the $WZ$+jets background. 
Nevertheless, in future it would be better to use $WZ$+jets control regions with simultaneous fits, as done to get the results in e.g the one lepton signal regions discussed in Ref.~\citen{ATLAS:2021fbt} (even if it is more time consuming).

In the selected UDD RPV discovery signal regions, the main uncertainties are the ones associated to the detector backgrounds -- not very surprising, as these have the electron charge flip and the fake/non-prompt leptons as the main background.
With more data available (possible with the LHC Run 3), these uncertainties can be reduced as the various efficiencies needed as input by the data-based methods can be measured more precisely.
Also the matrix method could be expanded to use as input efficiencies measured per fake/non-prompt lepton source.

In the bRPV signal regions, the total uncertainty is around 26\% in \SRtwolbrpv\ with ``Fakes/non-prompt" being the main contribution, while in \SRthreelbrpv\ it is around 16\% with the ``Fakes/non-prompt" and ``Modelling" categories contributing equally.

\section{Results}
\begin{figure}[!tb]
	\begin{center}
		\subfigure{\hspace{-0.cm}
			\includegraphics[width=0.5\columnwidth]{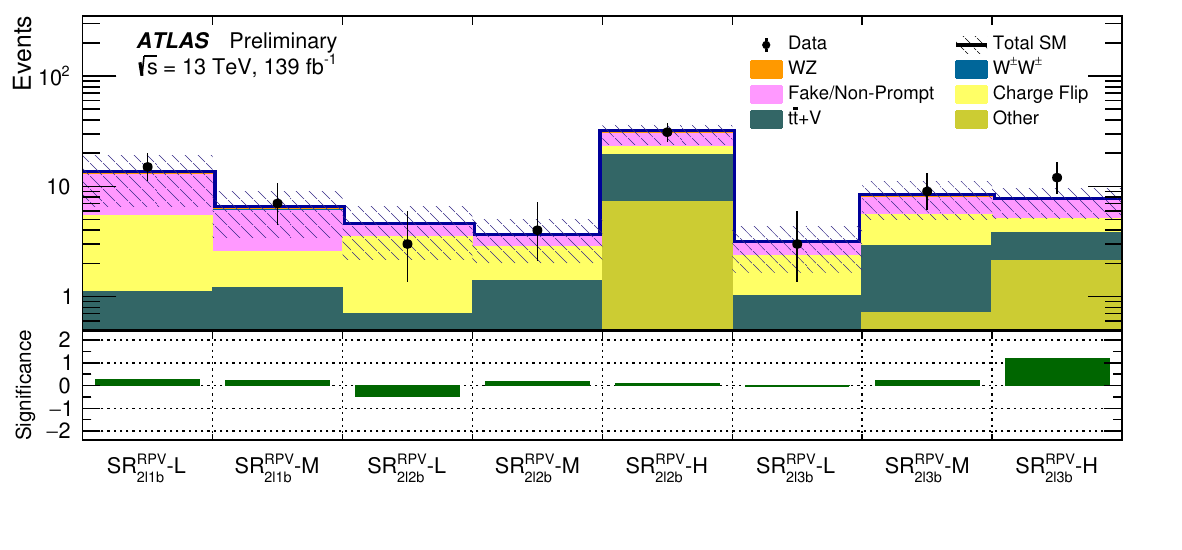}
		}
	\end{center}\vspace{-0.3cm}
	\caption{
		Results in the UDD RPV discovery signal regions. Reused with permission from Ref.~\citen{ATLAS-CONF-2022-057}. 
	}
	\label{fig:UDDRPV_SRs_yields}
\end{figure}

\begin{table}[!tb]
	\centering
	\tbl{
		95\% CL upper limits on the visible cross section times efficiency (\sigmavis). The upper limits on the observed signal events (\sobs), as well on the signal events given the expected number of background events (\sexp) and $\pm1\sigma$ variations of the expected number) are also shown. Reused with permission from Ref.~\citen{ATLAS-CONF-2022-057}.
	}
	{	\begin{tabular}{l|c|c|c}
			\toprule
			Signal region  & \sigmavis\ [fb]  &  \sobs  &  \sexp  \\
			\colrule
			\SRonebjrpv-L    & $0.13$ &  $17.5$ & $ { 15.1 }^{ +4.8 }_{ -3.7 }$ \\%
			\SRonebjrpv-M    & $0.07$ &  $10.1$ & $ { 8.9 }^{ +3.1 }_{ -1.7 }$ \\ \hline
			\SRtwobjrpv-L    & $0.04$ &  $6.1$ & $ { 6.2 }^{ +2.4 }_{ -1.1 }$  \\%
			\SRtwobjrpv-M    & $0.05$ &  $6.8$ & $ { 6.0 }^{ +2.3 }_{ -1.2 }$  \\%
			\SRtwobjrpv-H    & $0.15$ &  $20.7$ & $ { 18.6 }^{+6.0 }_{ -4.3 }$  \\ \hline
			\SRthreebjrpv-L    & $0.04$ &  $6.1$ & $ { 5.7 }^{ +1.9 }_{ -1.0 }$  \\%
			\SRthreebjrpv-M    & $0.08$ &  $11.5$ & $ { 9.7 }^{ +3.2 }_{ -1.8 }$  \\%
			\SRthreebjrpv-H    & $0.10$ &  $13.5$ & $ { 8.6 }^{ +3.2 }_{ -2.5 }$  \\%
			\botrule
	\end{tabular}}
	\label{tab:UDDRPV_MIUL}
\end{table}

As hinted by Figures~\ref{fig:UDDRPV_Nmin1_set2} and~\ref{fig:UDDRPV_Nmin1_set3} and summarized in Figure~\ref{fig:UDDRPV_SRs_yields}, there is no significant excess in any of the selected discovery signal regions (nor in the exclusion UDD RPV signal regions~\cite{ATLAS:2021fbt}). 
The highest excess is only around $1 \sigma$ in \SRthreebjrpv, and 
if the MC-Template method would be used instead of the data-based methods for the detector background, this excess would be even lower.
Using these yields, 95\% CL model independent upper limits on the number of observed (obs) and expected (exp) BSM events (S$^{95}$) that may contribute to the discovery signal regions are set.
Normalizing these to the luminosity of the data sample (139~fb$^{-1}$), upper limits on the visible BSM cross-section are also obtained: $\sigma$~=~$\sigma_{\mathrm{prod}}\times A\times \epsilon$~=~S$^{95}$/139~fb$^{-1}$.
$A$ and $\epsilon$ are the corresponding fiducial acceptance and selection efficiency of a BSM signal, in the considered signal region.
Selected results are presented in Table~\ref{tab:UDDRPV_MIUL}.

Model dependent exclusion limits are set on the UDD RPV $pp \to \chinoonepm \ninoone_{,2}, \; \ninoone\ninotwo$ production cross-section versus the $\ninoone$ (LSP) mass, and shown in Ref.~\citen{ATLAS-CONF-2022-057,ATLAS:2021fbt}.
Upper limits on the production cross section range from 6.5~pb to 0.18~pb, when the higgsino LSP mass varies from 180~GeV to 400~GeV.
LSP masses between 200~GeV and 320~GeV are excluded, thanks to the usage of a neural network in the signal regions defined with a one lepton selection, and to the $m^{\ell j} < 155$~GeV requirement applied for the two same-charge lepton selection. 
This was seen when performing the optimization of the signal regions in Ref.~\citen{ATLAS:2021fbt}, and confirmed in Ref.~\citen{ATLAS-CONF-2022-057} where the discovery signal regions from Table~\ref{tab:SR_UDDRPV} are used to obtain exclusion limits and only the 200~GeV mass point is excluded.
The conclusion is that, for discovery, more inclusive (general) signal regions should still be used, as in nature SUSY will not manifest exactly as in a simplified model, while for exclusion limits the usage of machine learning techniques will greatly help. 

\begin{figure}[!tb]
	\begin{center}
		\subfigure{\hspace{-1.3cm}
			\includegraphics[width=0.28\columnwidth]{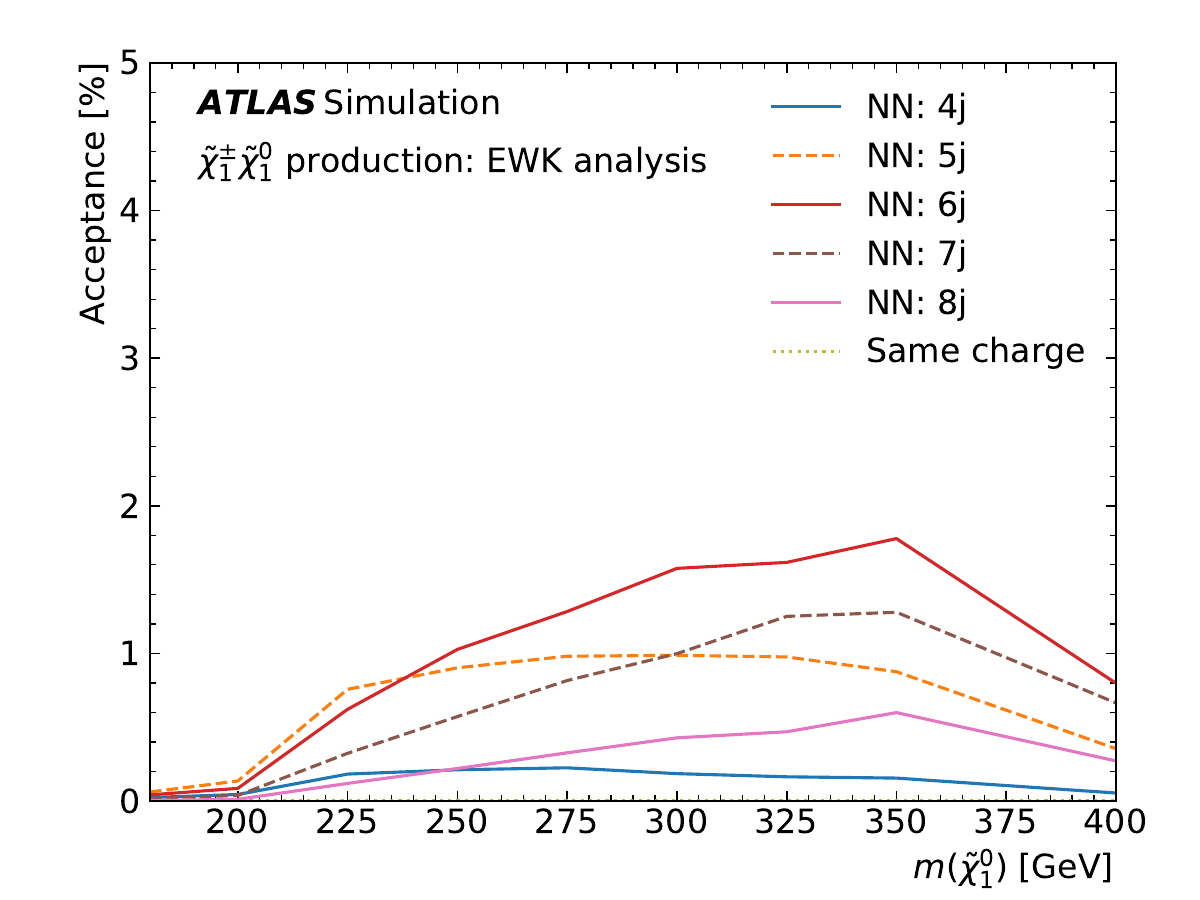}
		}\hspace{-0.7cm}	
		\subfigure{
			\includegraphics[width=0.28\columnwidth]{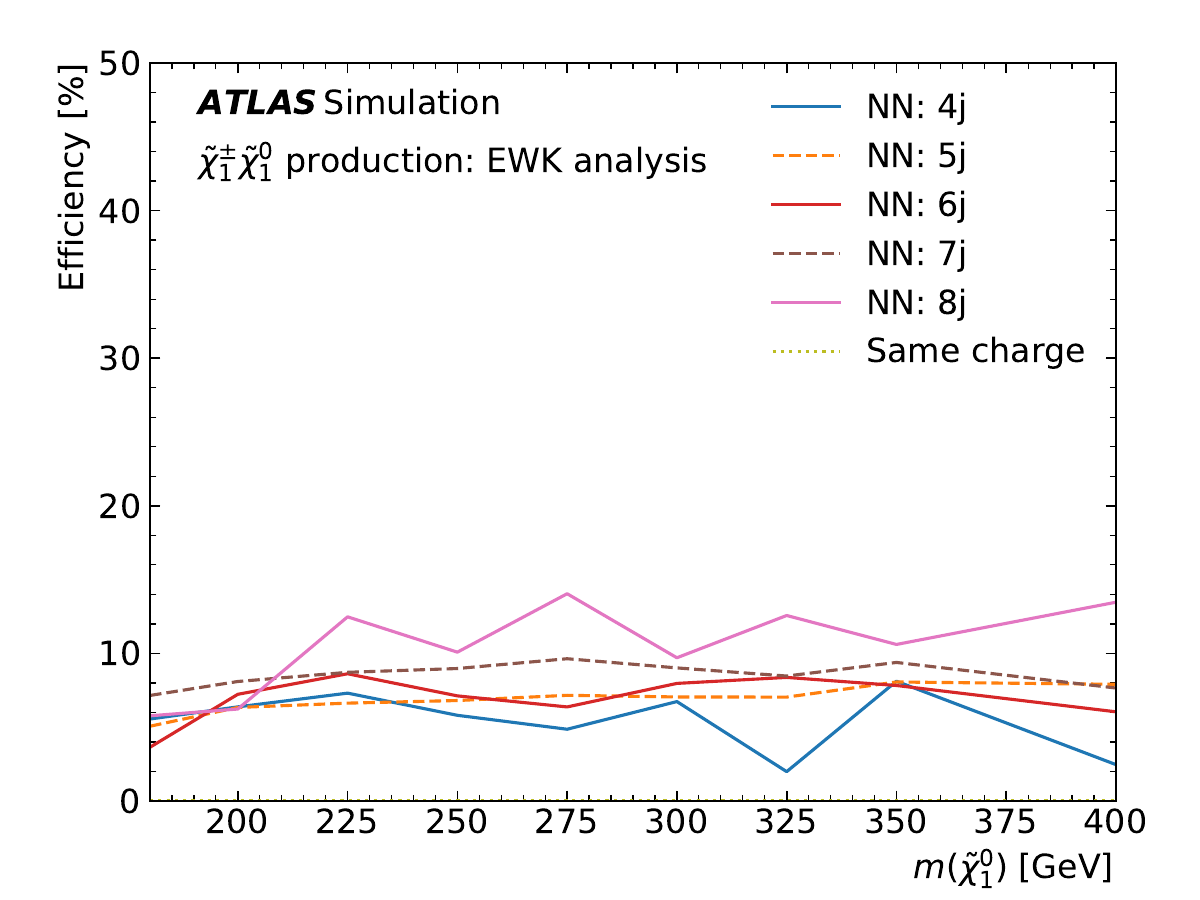}
		}\hspace{-0.7cm}	
		\subfigure{
			\includegraphics[width=0.28\columnwidth]{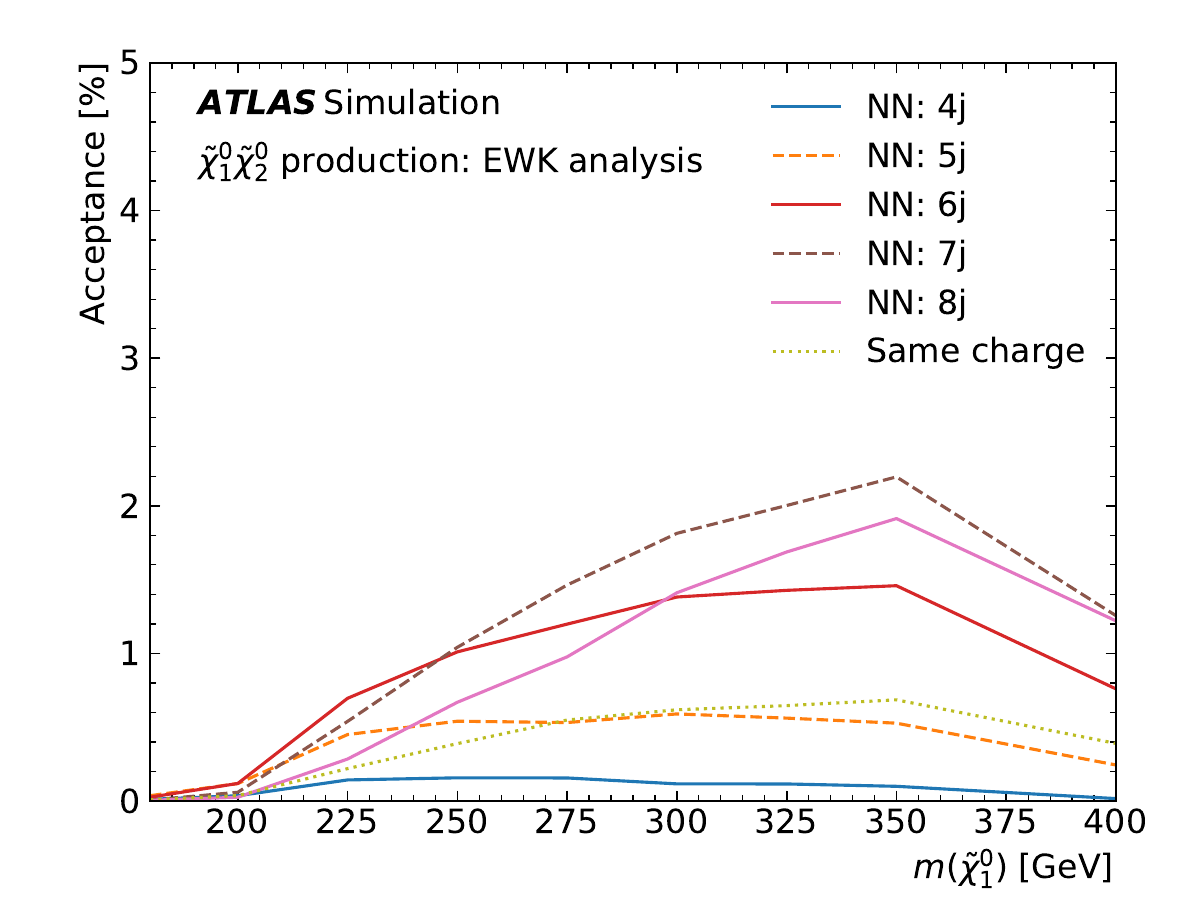}
		}\hspace{-0.7cm}	
		\subfigure{
			\includegraphics[width=0.28\columnwidth]{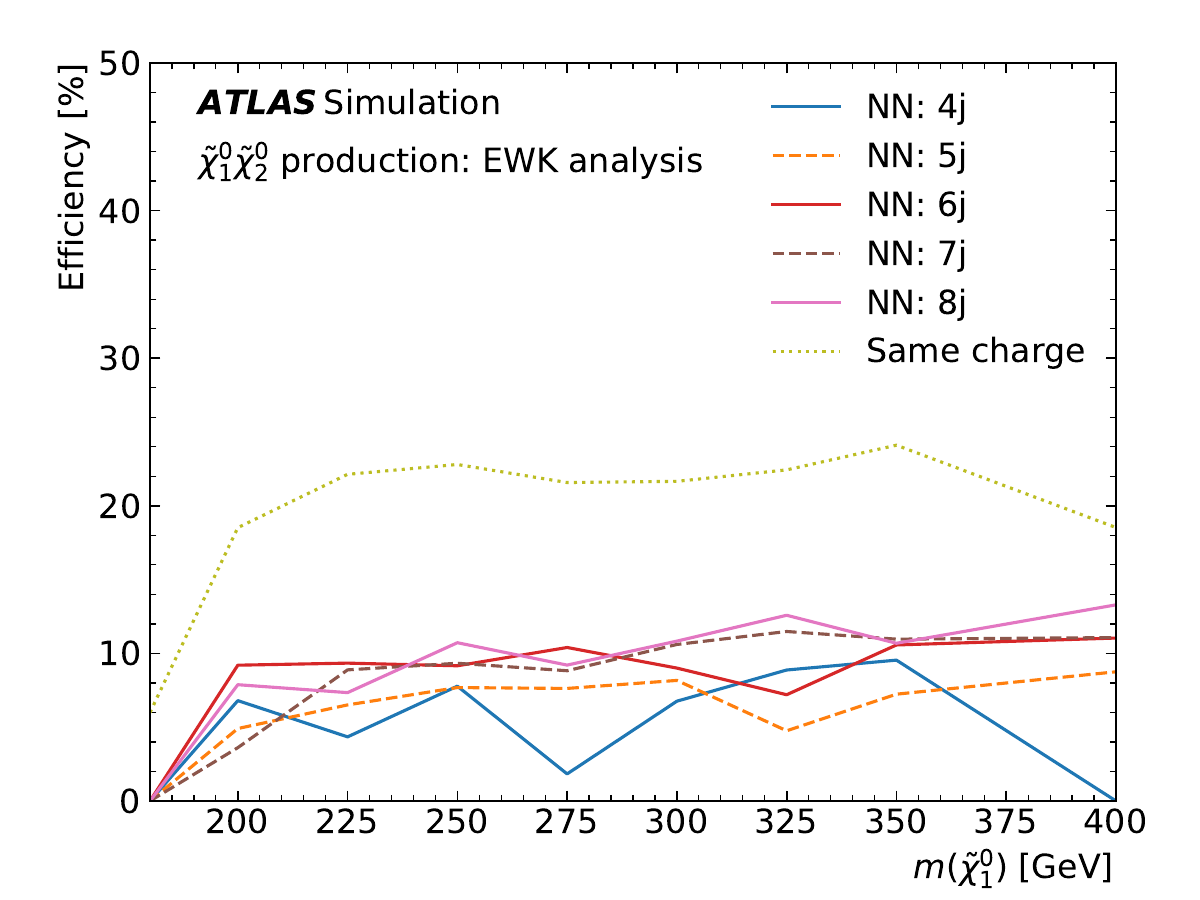}
		}
	\end{center}\vspace{-0.3cm}
	\caption{
		$\chinoonepm \ninoone$ (left) and $\ninoone\ninotwo$ (right) UDD RPV signal acceptance and efficiency. Reused with permission from Ref.~\citen{ATLAS-PPRPV1Lep}.
	}
	\label{fig:UDDRPV_Acc_Eff}
\end{figure}

Figure~\ref{fig:UDDRPV_Acc_Eff} shows the signal acceptance and efficiency for the two RPV UUD production modes illustrated in Figures~\ref{fig:diag_RPV1} and~\ref{fig:diag_RPV2}. 
The results are shown in the one lepton signal regions defined with requirements on the NN discriminant in the 4--8 jet bins, as well as in the two same-charge lepton signal {regions~\cite{ATLAS:2021fbt}}.
The acceptance gives the number of events passing the selection cuts at generator level,
and the efficiency accounts for reconstruction losses like lepton identification, jet energy resolution, jet tagging efficiency, \met\
resolution, etc.
For the $\chinoonepm \ninoone$ production mode, values of $A \times \epsilon$ are at maximum 15\%, while for the $\ninoone\ninotwo$ production mode they are 25\% in the one lepton signal regions, and 15\% in the two same-charge lepton signal regions, respectively.

With the bRPV model, the $\ninoone$, $\ninotwo$, $\chinopm$ masses are excluded up to 440~GeV, assuming an inclusive higgsino production
and allowing all predicted sparticle decay modes.
The upper limits on the production cross section range from 0.18~pb to 0.025~pb, when the higgsino masses vary from 200~GeV to 700~GeV.
In $\SRtwolbrpv$ ($\SRthreelbrpv$), values of $A \times \epsilon$ are maximum of 1.5\% (5\%).

\vspace{-0.5cm}
\section{Conclusions}	
Two searches for RPV SUSY through the direct production of pairs of higgsinos have been discussed.
The results have been obtained in final states with one, two same-charge or three leptons, using 139~fb$^{-1}$ of LHC data.
The analyses have been optimized using two RPV SUSY models, with the $R$-parity violation obtained though lepton number violation  or through baryon number violation. 
Only prompt decays have been considered.
The methods used to estimate the background in two same-charge and three lepton regions have been discussed in detail, as well as their shortcomings. 
Some ideas for improvements have also been mentioned. 
Finally, the model dependent and independent upper limits have been presented.

\vspace{0.7cm}\hspace{-1.cm}
Copyright 2023 CERN for the benefit of the ATLAS Collaboration. CC-BY-4.0 license.
\vspace{0.7cm}




\end{document}

%% file: mydefs.tex
\newcommand{\ttZ}{\ensuremath{t\bar t + Z}}
\newcommand{\ttW}{\ensuremath{t\bar t + W}}
\newcommand{\ttV}{\ensuremath{t\bar t + V}}
\newcommand{\WZ}{\ensuremath{WZ}}

\newcommand{\nJ}{\ensuremath{n_{\mathrm{jets}}}\xspace}
\newcommand{\nbJ}{\ensuremath{n_{b-\mathrm{jets}}}\xspace}

\newcommand{\pT}{\ensuremath{p_T}\xspace}

\newcommand{\met}{\ensuremath{E_\mathrm{T}^\mathrm{miss}}\xspace}

\newcommand{\mttwo}{\ensuremath{m_\mathrm{T2}}\xspace}

\newcommand{\meff}{\ensuremath{m_\mathrm{eff}}\xspace}
\newcommand{\fracbjet}{\ensuremath{\sum\pT^{b-\mathrm{jet}}/\sum\pT^{\mathrm{jet}}}\xspace}
\newcommand{\DRlj}{\ensuremath{\Delta R (\ell_{1}, \mathrm{jet})_{\min}}\xspace}
\newcommand{\sumjet}{\ensuremath{\sum\pT^{\mathrm{jet}}}\xspace}

\def\antibar#1{\ensuremath{#1\bar{#1}}}

\def\ttbar{\antibar{t}}

\newcommand{\SRtwolbrpv}{\ensuremath{\mathrm{SR}^{\mathrm{bRPV}}_{2\ell}}\xspace}
\newcommand{\SRthreelbrpv}{\ensuremath{\mathrm{SR}^{\mathrm{bRPV}}_{3\ell}}\xspace}
\newcommand{\SRonebjrpv}{\ensuremath{\mathrm{SR}^{\mathrm{RPV}}_{2\ell1b}}\xspace}
\newcommand{\SRtwobjrpv}{\ensuremath{\mathrm{SR}^{\mathrm{RPV}}_{2\ell2b}}\xspace}
\newcommand{\SRthreebjrpv}{\ensuremath{\mathrm{SR}^{\mathrm{RPV}}_{2\ell3b}}\xspace}
\newcommand{\SRrpv}{\ensuremath{\mathrm{SR}^{\mathrm{RPV}}}\xspace}

\newcommand{\CRWZtwoj}{\ensuremath{\mathrm{CR}\WZ^{\WZ,\mathrm{(b)RPV}}_{2j}}\xspace}

\newcommand{\VRWZfourj}{\ensuremath{\mathrm{VR}\WZ^{\WZ,\mathrm{(b)RPV}}_{4j}}\xspace}
\newcommand{\VRWZfivej}{\ensuremath{\mathrm{VR}\WZ^{\WZ,\mathrm{(b)RPV}}_{5j}}\xspace}
\newcommand{\VRttV}{\ensuremath{\mathrm{VR}\ttV^{\WZ,\mathrm{(b)RPV}}}\xspace}

\newcommand{\sigmavis}{\ensuremath{\langle\epsilon\sigma\rangle_{\mathrm{obs}}^{95}}\xspace}
\newcommand{\sobs}{\ensuremath{S_\mathrm{obs}^{95}}\xspace}
\newcommand{\sexp}{\ensuremath{S_\mathrm{exp}^{95}}\xspace}

\def\susy#1{\ensuremath{\tilde{#1}}}%
\def\ggino{\ensuremath{\mathchoice%
      {\displaystyle\raise.4ex\hbox{$\displaystyle\tilde\chi$}}%
         {\textstyle\raise.4ex\hbox{$\textstyle\tilde\chi$}}%
       {\scriptstyle\raise.3ex\hbox{$\scriptstyle\tilde\chi$}}%
 {\scriptscriptstyle\raise.3ex\hbox{$\scriptscriptstyle\tilde\chi$}}}}

\def\chinop{\ensuremath{\mathchoice%
      {\displaystyle\raise.4ex\hbox{$\displaystyle\tilde\chi^+$}}%
         {\textstyle\raise.4ex\hbox{$\textstyle\tilde\chi^+$}}%
       {\scriptstyle\raise.3ex\hbox{$\scriptstyle\tilde\chi^+$}}%
 {\scriptscriptstyle\raise.3ex\hbox{$\scriptscriptstyle\tilde\chi^+$}}}}
\def\chinom{\ensuremath{\mathchoice%
      {\displaystyle\raise.4ex\hbox{$\displaystyle\tilde\chi^-$}}%
         {\textstyle\raise.4ex\hbox{$\textstyle\tilde\chi^-$}}%
       {\scriptstyle\raise.3ex\hbox{$\scriptstyle\tilde\chi^-$}}%
 {\scriptscriptstyle\raise.3ex\hbox{$\scriptscriptstyle\tilde\chi^-$}}}}
\def\chinopm{\ensuremath{\mathchoice%
      {\displaystyle\raise.4ex\hbox{$\displaystyle\tilde\chi^\pm$}}%
         {\textstyle\raise.4ex\hbox{$\textstyle\tilde\chi^\pm$}}%
       {\scriptstyle\raise.3ex\hbox{$\scriptstyle\tilde\chi^\pm$}}%
 {\scriptscriptstyle\raise.3ex\hbox{$\scriptscriptstyle\tilde\chi^\pm$}}}}
\def\chinomp{\ensuremath{\mathchoice%
      {\displaystyle\raise.4ex\hbox{$\displaystyle\tilde\chi^\mp$}}%
         {\textstyle\raise.4ex\hbox{$\textstyle\tilde\chi^\mp$}}%
       {\scriptstyle\raise.3ex\hbox{$\scriptstyle\tilde\chi^\mp$}}%
 {\scriptscriptstyle\raise.3ex\hbox{$\scriptscriptstyle\tilde\chi^\mp$}}}}

\def\chinoonep{\ensuremath{\mathchoice%
      {\displaystyle\raise.4ex\hbox{$\displaystyle\tilde\chi^+_1$}}%
         {\textstyle\raise.4ex\hbox{$\textstyle\tilde\chi^+_1$}}%
       {\scriptstyle\raise.3ex\hbox{$\scriptstyle\tilde\chi^+_1$}}%
 {\scriptscriptstyle\raise.3ex\hbox{$\scriptscriptstyle\tilde\chi^+_1$}}}}
\def\chinoonem{\ensuremath{\mathchoice%
      {\displaystyle\raise.4ex\hbox{$\displaystyle\tilde\chi^-_1$}}%
         {\textstyle\raise.4ex\hbox{$\textstyle\tilde\chi^-_1$}}%
       {\scriptstyle\raise.3ex\hbox{$\scriptstyle\tilde\chi^-_1$}}%
 {\scriptscriptstyle\raise.3ex\hbox{$\scriptscriptstyle\tilde\chi^-_1$}}}}
\def\chinoonepm{\ensuremath{\mathchoice%
      {\displaystyle\raise.4ex\hbox{$\displaystyle\tilde\chi^\pm_1$}}%
         {\textstyle\raise.4ex\hbox{$\textstyle\tilde\chi^\pm_1$}}%
       {\scriptstyle\raise.3ex\hbox{$\scriptstyle\tilde\chi^\pm_1$}}%
 {\scriptscriptstyle\raise.3ex\hbox{$\scriptscriptstyle\tilde\chi^\pm_1$}}}}

\def\chinotwop{\ensuremath{\mathchoice%
      {\displaystyle\raise.4ex\hbox{$\displaystyle\tilde\chi^+_2$}}%
         {\textstyle\raise.4ex\hbox{$\textstyle\tilde\chi^+_2$}}%
       {\scriptstyle\raise.3ex\hbox{$\scriptstyle\tilde\chi^+_2$}}%
 {\scriptscriptstyle\raise.3ex\hbox{$\scriptscriptstyle\tilde\chi^+_2$}}}}
\def\chinotwom{\ensuremath{\mathchoice%
      {\displaystyle\raise.4ex\hbox{$\displaystyle\tilde\chi^-_2$}}%
         {\textstyle\raise.4ex\hbox{$\textstyle\tilde\chi^-_2$}}%
       {\scriptstyle\raise.3ex\hbox{$\scriptstyle\tilde\chi^-_2$}}%
 {\scriptscriptstyle\raise.3ex\hbox{$\scriptscriptstyle\tilde\chi^-_2$}}}}
\def\chinotwopm{\ensuremath{\mathchoice%
      {\displaystyle\raise.4ex\hbox{$\displaystyle\tilde\chi^\pm_2$}}%
         {\textstyle\raise.4ex\hbox{$\textstyle\tilde\chi^\pm_2$}}%
       {\scriptstyle\raise.3ex\hbox{$\scriptstyle\tilde\chi^\pm_2$}}%
 {\scriptscriptstyle\raise.3ex\hbox{$\scriptscriptstyle\tilde\chi^\pm_2$}}}}

\def\nino{\ensuremath{\mathchoice%
      {\displaystyle\raise.4ex\hbox{$\displaystyle\tilde\chi^0$}}%
         {\textstyle\raise.4ex\hbox{$\textstyle\tilde\chi^0$}}%
       {\scriptstyle\raise.3ex\hbox{$\scriptstyle\tilde\chi^0$}}%
 {\scriptscriptstyle\raise.3ex\hbox{$\scriptscriptstyle\tilde\chi^0$}}}}

\def\ninoone{\ensuremath{\mathchoice%
      {\displaystyle\raise.4ex\hbox{$\displaystyle\tilde\chi^0_1$}}%
         {\textstyle\raise.4ex\hbox{$\textstyle\tilde\chi^0_1$}}%
       {\scriptstyle\raise.3ex\hbox{$\scriptstyle\tilde\chi^0_1$}}%
 {\scriptscriptstyle\raise.3ex\hbox{$\scriptscriptstyle\tilde\chi^0_1$}}}}
\def\ninotwo{\ensuremath{\mathchoice%
      {\displaystyle\raise.4ex\hbox{$\displaystyle\tilde\chi^0_2$}}%
         {\textstyle\raise.4ex\hbox{$\textstyle\tilde\chi^0_2$}}%
       {\scriptstyle\raise.3ex\hbox{$\scriptstyle\tilde\chi^0_2$}}%
 {\scriptscriptstyle\raise.3ex\hbox{$\scriptscriptstyle\tilde\chi^0_2$}}}}
\def\ninothree{\ensuremath{\mathchoice%
      {\displaystyle\raise.4ex\hbox{$\displaystyle\tilde\chi^0_3$}}%
         {\textstyle\raise.4ex\hbox{$\textstyle\tilde\chi^0_3$}}%
       {\scriptstyle\raise.3ex\hbox{$\scriptstyle\tilde\chi^0_3$}}%
 {\scriptscriptstyle\raise.3ex\hbox{$\scriptscriptstyle\tilde\chi^0_3$}}}}
\def\ninofour{\ensuremath{\mathchoice%
      {\displaystyle\raise.4ex\hbox{$\displaystyle\tilde\chi^0_4$}}%
         {\textstyle\raise.4ex\hbox{$\textstyle\tilde\chi^0_4$}}%
       {\scriptstyle\raise.3ex\hbox{$\scriptstyle\tilde\chi^0_4$}}%
 {\scriptscriptstyle\raise.3ex\hbox{$\scriptscriptstyle\tilde\chi^0_4$}}}}








%% file: RPV_SUSY_EWK_2023.bbl
\vspace{-0.5cm}
\begin{thebibliography}{99}\vspace{-0.cm}
\bibitem{Martin:1997ns}	S.~P.~Martin, \emph{A Supersymmetry primer}, Adv. Ser. Direct. High Energy Phys. \textbf{18} (1998), 1-98, [arXiv:hep-ph/9709356 [hep-ph]]
\bibitem{Evans:2008zzb} L.~Evans and P.~Bryant, \emph{LHC Machine}, JINST {\bf 3} (2008) S08001 
\bibitem{Dreiner:1997uz} H.~K.~Dreiner, \emph{An Introduction to explicit R-parity violation}, Adv. Ser. Direct. High Energy Phys. \textbf{21} (2010), 565-583, [arXiv:hep-ph/9707435 [hep-ph]]
\bibitem{ATLAS-CONF-2022-057} ATLAS Collaboration, \emph{Search for direct production of winos and higgsinos in events with two same-sign or three leptons in pp collision data at 13 TeV with the ATLAS detector}, JHEP 11 (2023) 150, [2305.09322 [hep-ex]]
\bibitem{ATLAS:2021fbt} ATLAS Collaboration, \emph{Search for R-parity-violating supersymmetry in a final state containing leptons and many jets with the ATLAS experiment using $\sqrt{s} = 13 { TeV}$ proton\textendash{}proton collision data}, Eur. Phys. J. C \textbf{81} (2021) no.11, 1023, [arXiv:2106.09609 [hep-ex]]
\bibitem{SUSY_ATLAS} ATLAS collaboration, A list of the published ATLAS Supersymmetry searches,  \link{https://twiki.cern.ch/twiki/bin/view/AtlasPublic/SupersymmetryPublicResults}{twiki.cern.ch/AtlasPublic/SupersymmetryPublicResults}
\bibitem{SUSY_CMS} CMS collaboration, A list of the published CMS Supersymmetry searches, \link{https://twiki.cern.ch/twiki/bin/view/CMSPublic/PhysicsResultsSUS}{twiki.cern.ch/CMSPublic/PhysicsResultsSUSY}
\bibitem{Atlas_Detector} ATLAS Collaboration, \emph{The ATLAS Experiment at the CERN Large Hadron Collider}, JINST {\bf 3} (2008) S08003 
\bibitem{ATLAS:2016wtr} ATLAS Collaboration, \emph{Performance of the ATLAS Trigger System in 2015}, Eur. Phys. J. C \textbf{77} (2017) no.5, 317, [arXiv:1611.09661 [hep-ex]]
\bibitem{ATLAS:2019dpa} ATLAS Collaboration, \emph{Performance of electron and photon triggers in ATLAS during LHC Run 2}, Eur. Phys. J. C \textbf{80} (2020) no.1, 47, [arXiv:1909.00761 [hep-ex]]
\bibitem{ATLAS:2020gty} ATLAS Collaboration, \emph{Performance of the ATLAS muon triggers in Run 2}, JINST \textbf{15} (2020) no.09, P09015, [arXiv:2004.13447 [physics.ins-det]]
\bibitem{ATLAS:2020atr} ATLAS Collaboration, \emph{Performance of the missing transverse momentum triggers for the ATLAS detector during Run-2 data taking}, JHEP \textbf{08} (2020), 080, [arXiv:2005.09554 [hep-ex]]
\bibitem{ATL-SOFT-PUB-2021-001} ATLAS Collaboration, \emph{The ATLAS Collaboration Software and Firmware}, ATL-SOFT-PUB-2021-001, \url{https://cds.cern.ch/record/2767187}
\bibitem{Baak:2014wma} M.~Baak, G.~J.~Besjes, D.~C\^ote, A.~Koutsman, J.~Lorenz and D.~Short, \emph{HistFitter software framework for statistical data analysis}, Eur. Phys. J. C \textbf{75} (2015), 153, [arXiv:1410.1280 [hep-ex]]
\bibitem{ATLAS:2017ghe} ATLAS Collaboration, \emph{Jet reconstruction and performance using particle flow with the ATLAS Detector}, Eur. Phys. J. C \textbf{77} (2017) no.7, 466, [arXiv:1703.10485 [hep-ex]]
\bibitem{ATLAS:2019bwq} ATLAS Collaboration, \emph{ATLAS b-jet identification performance and efficiency measurement with $t{\bar{t}}$ events in pp collisions at $\sqrt{s}=13$ TeV}, Eur. Phys. J. C \textbf{79} (2019) no.11, 970, [arXiv:1907.05120 [hep-ex]]
\bibitem{ATLAS:bjets} ATLAS Collaboration, \emph{Optimisation and performance studies of the ATLAS b-tagging algorithms for the 2017-18 LHC run}, ATL-PHYS-PUB-2017-013, \url{https://cds.cern.ch/record/2273281}
\bibitem{ATLAS-PPRPV1Lep} ATLAS Collaboration,\emph{Search for R-parity-violating supersymmetry in a final state containing leptons and many jets with the ATLAS experiment using proton proton collision data} -- the public page, \url{https://atlas.web.cern.ch/Atlas/GROUPS/PHYSICS/PAPERS/SUSY-2019-04/}
\bibitem{ATLAS:2019bsc} ATLAS Collaboration, \emph{Measurement of $W^{\pm}Z$ production cross sections and gauge boson polarisation in $pp$ collisions at $\sqrt{s} = 13$ TeV with the ATLAS detector}, Eur. Phys. J. C \textbf{79} (2019) no.6, 535, [arXiv:1902.05759 [hep-ex]]
\bibitem{ATLAS:2019qmc} ATLAS Collaboration, \emph{Electron and photon performance measurements with the ATLAS detector using the 2015\textendash{}2017 LHC proton-proton collision data}, JINST \textbf{14} (2019) no.12, P12006, [arXiv:1908.00005 [hep-ex]]
\bibitem{ATLAS_ChFlipBDT} ATLAS Collaboration, \emph{Search for direct production of winos and higgsinos in events with two same-sign or three leptons in pp collision data at 13 TeV with the ATLAS detector} -- the public page,  \url{https://atlas.web.cern.ch/Atlas/GROUPS/PHYSICS/PAPERS/SUSY-2016-14/}
\bibitem{ATLAS:2022swp} ATLAS Collaboration, \emph{Tools for estimating fake/non-prompt lepton backgrounds with the ATLAS detector at the LHC}, submitted to JINST, [arXiv:2211.16178 [hep-ex]]
\bibitem{ATLAS:2019fag} ATLAS Collaboration, \emph{Search for squarks and gluinos in final states with same-sign leptons and jets using 139 fb$^{-1}$ of data collected with the ATLAS detector}, JHEP \textbf{06} (2020), 046, [arXiv:1909.08457 [hep-ex]]
\end{thebibliography}
